\documentclass[12pt]{article}
\usepackage{amsmath,amssymb,authblk,graphicx}

\newcommand{\Hom}{{\rm Hom}}
\newcommand{\ra}{{\rightarrow}}

\newcommand{\ZZ}{{\mathbb Z}}
\newcommand{\RR}{{\mathbb R}}
\newcommand{\CC}{{\mathbb C}}

\newcommand{\Vect}{{\rm Vect}}
\newcommand{\Cob}{{\rm Cob}}
\newcommand{\cA}{{\mathcal A}}
\newcommand{\Arf}{{\rm Arf}}
\newcommand{\cC}{{\mathcal C}}
\newcommand{\cD}{{\mathcal D}}
\newcommand{\sVect}{{\rm sVect}}
\newcommand{\kk}{{\mathbf k}}
\newcommand{\Ob}{{\rm Ob}}
\newcommand{\Cl}{{\rm Cl}}
\renewcommand{\aa}{{\mathbf a}}
\newcommand{\BI}{{\bar i}}
\newcommand{\bj}{{\bar j}}
\newcommand{\bk}{{\bar k}}
\newcommand{\eps}{\epsilon}
\newcommand{\Fj}[6]{F\begin{bmatrix} #1 & #2 & #3 \\ #4 & #5 & #6\end{bmatrix}}
\newcommand{\bone}{{\bf 1}}

\begin{document}

\title{Spin TQFTs and fermionic phases of matter}

\author{Davide Gaiotto} 
\affil{{\it Perimeter Institute for Theoretical Physics,} \\ \it Waterloo, Ontario, Canada N2L 2Y5}
\author{Anton Kapustin}
\affil{{\it Simons Center for Geometry and Physics,} \\ \it Stony Brook, NY 11790}

\maketitle

\abstract{We study lattice constructions of gapped fermionic phases of matter. We
show that the construction of fermionic Symmetry Protected Topological
orders by Gu and Wen has a hidden dependence on a discrete spin
structure on the Euclidean space-time.  The spin structure is needed to
resolve ambiguities which are otherwise present. An identical ambiguity is
shown to arise in the fermionic analog of the string-net construction of
2D topological orders. We argue that the need for a spin structure is a
general feature of lattice models with local fermionic degrees of freedom and is
a lattice analog of the spin-statistics relation.}

\section{Introduction and summary}

\subsection{Bosonic and fermionic gapped phases}
In condensed matter physics, topological phases of matter are often defined as equivalence classes of local gapped bosonic Hamiltonians, 
usually defined on a lattice, which can be deformed into each other without ever becoming gapless \cite{ChenGuWen,Kitaevtalk}. The notion of topological phase can 
be enriched by imposing additional constraints on the theories, such as a choice of global symmetry preserved by all the Hamiltonians. 
On the other hand, topological quantum field theories\footnote{This is a somewhat looser notion of TQFT compared to some formal definitions. For example, 
we consider Chern-Simons theory to be a TQFT, even though it has a partition function which depends on a choice of metric on space-time. 
In other words, we allow the stress tensor to be non-zero, but proportional to the identity operator.} can be thought of as describing the far infrared behavior of 
gapped bosonic quantum field theories (see e.g. section 4 of \cite{Witten:index} or the monograph \cite{Wen:mono}).

There is a close relation between topological phases of matter and topological quantum field theories, which can be thought of as a 
map from a topological phase of matter to the TQFT which encodes the low energy continuum limit of the corresponding Hamiltonian. 
In principle, one may imagine the map being many-to-one: it is not obvious that two local gapped Hamiltonians 
which map to the same TQFT will always be deformable into each other. Still, in practice we do not know of any observable 
which can distinguish two phases of matter, but cannot be formulated in terms of the TQFT data. \footnote{It is also conceivable, perhaps, that some 
topological phase of matter may not give rise to a TQFT at low energy, i.e. that some anomaly/obstruction may prevent the definition of 
TQFT amplitudes on general manifolds in terms of the Hamiltonian data. But in all cases known to us one circumvent such obstructions by postulating that the 
TQFT depends on some additional geometric data, such as metric or framing.}

In condensed matter physics, one also encounters the notion of a fermionic topological phase of matter, 
defined as an equivalence class of local gapped Hamiltonians which can involve fermionic 
degrees of freedom \cite{ChenGuWen2,KitaevFidkowski}. Perhaps surprisingly, some fermionic phases of matter are not expected to admit a purely bosonic realization. 
This is expected to be due to the difference in the notion of locality for bosonic and fermionic systems. Intuitively, 
if we partition a bosonic system in two parts, the total Hilbert space factors uniquely in the tensor product of the 
Hilbert spaces for the two parts. If we partition a fermionic system, though, the factorization has an intrinsic ambiguity, 
as observables in the tensor product of the Hilbert spaces for the two parts are defined up to a sign in the sector 
where both factors have odd fermion number. 

\subsection{Spin structure dependence}

In unitary quantum field theory, fermions are naturally spinors and thus the low energy physics of a gapped fermionic theory 
is a spin-TQFT: a topological field theory defined on manifolds which can be equipped with a spin structure, 
whose correlation functions possibly depend on the choice of spin structure. 
The purpose of this paper is to explore the relation between fermionic topological phases of matter 
and spin-TQFTs. It is not obvious that such a relation should exist, as a lattice Hamiltonian involving fermionic degrees of freedom 
is usually written down without any reference to a spin structure on the manifold which is discretized by the lattice. 
One also cannot appeal to the spin-statistics relation, because the lattice destroys Lorenz and even rotational invariance which are the conditions of the spin-statistics theorem. 
\footnote{Taking the continuum limit and then applying the spin-statistics relation does not ameliorate the problem. The TQFT itself is, of course, Lorentz invariant, but
the spin-statistics relation is a property of Lorenz invariant particle excitations. The continuum limit from the lattice theory to the low-energy
TQFT only concerns the ground states of the system. A priori, massive excitations above these ground states do not need to transform properly under the Lorentz group. }

The first step of our analysis is to look carefully at the fermionic SPT phases constructed by Gu and Wen in \cite{GuWen}.
We find that the prescription used to define the partition function of such theories runs into an obstruction if applied to space-time
manifolds of general topology, unless the second Stiefel-Whitney class $[w_2]$ of the manifolds vanishes, 
i.e. the manifold admits a spin structure. If the manifolds admits a spin structure, the obstruction can be eliminated, but the 
final answer will depend on the choice of spin structure $\eta$. In other words, these fermionic SPT phases
define (invertible) spin-TQFTs. 

Next, we look at other known constructions of fermionic phases of matter which are expected to admit a
state-sum-like definition of their partition function: the construction of fermionic toric code in \cite{fermionictoriccode} 
and the general fermionic Turaev-Viro construction in \cite{fermionicTV}. These references focus on the construction 
of a fixed-point Hamiltonian and wave-function for these fermionic phases of matter, rather than a partition function. 
It is straightforward, though, to assemble the same ingredients into a partition sum, borrowing some ideas from the Gu-Wen
fermionic SPT phase construction. Again, we find an obstruction to define the partition sum unless the space-time manifolds admits a spin structure,
in which case one can remove the obstruction and define a well defined partition function which depends on the choice of spin structure $\eta$.
Thus these fermionic phases of matter are associated to spin-TQFTs.
 
We can describe the obstruction schematically here, referring the reader to sections \ref{sec:guwen} and \ref{sec:tv} for further details.
State-sum models assemble the partition function from 
a triangulation of the space-time manifold $X$: each simplex is associated to some tensor in the tensor product of 
vector spaces associated to the faces and the legs of these tensors are contracted together as the simplices are glued along the corresponding
faces of the triangulation. In a fermionic model, the vector spaces may be Grassmann-odd and Koszul signs 
occur when re-organizing and contracting the factors of the tensor products. 

These Koszul signs, arising from the anti-commutation of fermionic variables, are of course a key element of the problem.
The non-local nature of these signs is precisely what should allow these fermionic phases of matter to be distinct from any bosonic phase.   
In order for the partition sum to be invariant under local changes in the triangulation of the manifold, 
one needs to cancel the change in the Koszul signs agains the change in the local data attached to the simplices. 
The obstruction arises precisely when this cancellation is not possible. 

We can express the obstruction neatly by encoding 
the fermion number of the vector spaces attached to faces in a $\ZZ_2$-valued 
$(d-1)$-cochain $\beta_{d-1}$. The cochain $\beta_{d-1}$ is actually a cocycle, 
as the total fermion number of the tensors attached to simplices 
is even. It is useful to decompose the partition sum into a sum 
of terms $Z[X,\beta_{d-1}]$, which contain the parts of the state sum due to states of fermion number $\beta_{d-1}$. 

We can encode a general change of triangulation of $X$ into a triangulation of the $(d+1)$-dimensional manifold 
$X \times [0,1]$. Intuitively, we are gluing a sequence of $(d+1)$-dimensional simplices on top of our initial triangulation to get the 
final triangulation. We find that the triangulation invariance of the partition function is obstructed by 
some irreducible sign mismatch, which can be written schematically as 
\begin{equation} \label{eq:obstruct}
(-1)^{\int_{X \times [0,1]} w_2 \cup \beta_{d-1}}
\end{equation} 
Here $w_2$ is a 2-cocycle with values in $\ZZ_2$ representing the second Stiefel-Whitney class of $X \times [0,1]$ and $\beta_{d-1}$ 
is a lift to $X \times [0,1]$ of the cocycle $\beta_{d-1}$. 
If the cohomology class of $[w_2]$ is non-trivial and the theory involves choices of fermion numbers $\beta_{d-1}$ which are non-trivial in cohomology, 
this sign mismatch cannot be absorbed by a redefinition of the local part of the partition function. 
This prevents us from constructing a well-defined partition sum and ruins the state-sum construction. 

If we restrict $X$ to be a spin manifold then $w_2(X)$ is exact and we can write $w_2=\delta\eta$ for some 1-cochain $\eta$,
which represents a choice of spin structure. This allows us to  
thus cancel the obstruction \ref{eq:obstruct} by the variation of a local term 
\begin{equation} \label{eq:counter}
(-1)^{\int_{X} \eta \cup \beta_{d-1}}
\end{equation} 
so that the improved state sum 
\begin{equation}
Z[X,\eta] = \sum_{\beta_{d-1}} Z[X,\beta_{d-1}] (-1)^{\int_{X} \eta\cup \beta_{d-1}}
\end{equation}
is fully invariant under changes of triangulations and defines a good theory. This theory 
is a spin-TQFT: it can only be defined on a spin manifold and depends on a choice of spin structure.

In sections \ref{sec:quadratic} and \ref{sec:anomaly} we will look in further detail at the properties of the 
Koszul signs which occur in the state sum. The definition the partition function requires 
specific choices of how to order the factors in the tensor product associated to each simplex, and the two factors in 
the contraction of vector spaces at each face. Given some ordering choices $\Pi$, the permutations 
of the vector spaces involved in the state sum will produce some overall Koszul sign 
$\sigma_\Pi(X,\beta_{d-1})$, which depends only on the triangulation, on $\Pi$ and on $\beta_{d-1}$. 

The choice of order $\Pi$ can be given independently of the other data in the state sum. The combined sign
\begin{equation} z_\Pi[X,\eta,\beta_{d-1}]= \sigma_\Pi(X,\beta_{d-1})(-1)^{\int_{X} \eta\cup \beta_{d-1}}
\end{equation}
appears to be a very useful object, which captures the intrinsically fermionic part of the full partition function. 
From now on we will drop the subscript $\Pi$. Our formulae will refer to the specific choice of order used in the
Gu-Wen definition of fermionic SPT phases \cite{GuWen}. We will comment briefly on other choices of order 
in section \ref{sec:quadratic}.

We can think about $z[X,\eta,\beta_{d-1}]$ as defining an effective action for 
a $(d-1)$-form $\ZZ_2$ gauge field with a very specific anomaly, 
or a very simple invertible spin-TQFT $K_d$ equipped with an anomalous 
$(d-2)$-form $\ZZ_2$ global symmetry. 

Under changes of triangulation, $z[X,\eta,\beta_{d-1}]$ changes by another interesting cocycle, the Steenrod square of $\beta_{d-1}$:
\begin{equation}
(-1)^{ \int_{X \times [0,1]} Sq^2[\beta_{d-1}]} \equiv (-1)^{ \int_{X \times [0,1]} \beta_{d-1} \cup_{d-3} \beta_{d-1}}
\end{equation}
We refer to appendix \ref{sec:higher} for the explicit definition of the higher cup products $\cup_a$. Their basic property is
\begin{equation} \label{eq:highercup}
A \cup_a B + B \cup_a A = \delta( A \cup_{a+1} B) + \delta A \cup_{a+1} B + A \cup_{a+1} \delta B
\end{equation}
with $\cup_0 \equiv \cup$. 

Under gauge transformations, we find the precise form of the 't Hooft anomaly 
 \begin{equation}
z[X,\eta,\beta_{d-1}+ \delta \lambda_{d-2}] =z[X,\eta,\beta_{d-1}] (-1)^{\int_X \beta \cup_{d-3} \lambda + \lambda \cup_{d-3} \beta + \lambda \cup_{d-3} \delta  \lambda + \lambda \cup_{d-4} \lambda}
\end{equation}

Although $z[X,\eta,\beta_{d-1}]$ does not appear to admit a $d$-dimensional bosonic description, 
we also find that it is a quadratic refinement of a bosonic pairing: 
\begin{equation}
z[X,\eta,\beta_{d-1} + \beta'_{d-1}] = z[X,\eta,\beta_{d-1}]z[X,\eta,\beta'_{d-1}] (-1)^{ \int_{X} \beta_{d-1} \cup_{d-2}  \beta'_{d-1}}
\end{equation}
Finally, if $X$ is a boundary of a compact oriented $(d+1)$-manifold $Y$ and $d>2$, we find an explicit WZW-like expression for $z[X,\eta,\beta_{d-1}]$:
\begin{equation}
z[X,\eta,\beta_{d-1}] = (-1)^{  \int_{X} \eta\cup \beta_{b-1} + \int_Y Sq^2[\beta_{d-1}] + w_2 \cup \beta_{d-1} }
\end{equation}
Here we use the fact that for $d>2$ the cocycle $\beta_{d-1}$ can be extended to $Y$. The action is independent of the choice of $Y$ or of the way $\beta_{d-1}$ is extended from $X$ too $Y$ because the expression $Sq^2[\beta_{d-1}] + w_2 \cup \beta_{d-1}$ is exact for closed oriented $Y$ . This formula is particularly useful for $d=3$, since any closed oriented 3-manifold $X$ is a boundary of a compact oriented 4-manifold $Y$. 

With a bit of extra work, we can rewrite the partition function $Z[X,\eta]$ of our spin-TQFT as the partition function of a 
$(d-1)$-form $\ZZ_2$ gauge theory
\begin{equation}
Z[X,\eta] = \sum_{\beta_{d-1}} \tilde Z[X,\beta_{d-1}] z[X,\eta,\beta_{d-1}]
\end{equation}
where the gauge fields are coupled to two two sets of degrees of freedom: 
a standard bosonic TQFT equipped with a $(d-2)$-form $\ZZ_2$ global symmetry 
and partition function $\tilde Z[X,\beta_{d-1}]$, and the spin-TQFT $K_d$. 
The bosonic theory associated to $\tilde Z[X,\beta_{d-1}]$ must have a 't Hooft anomaly 
which cancels the 't Hooft anomaly of $K_d$, controlled by $Sq^2[\beta_{d-1}]$.

In order to make contact with concepts which are more familiar in condensed matter physics, 
it is useful to replace the notion of a TQFT with an anomalous global symmetry with the 
notion of a gapped boundary condition for a $(d+1)$-dimensional SPT phase,
protected by a $(d-2)$-form $\ZZ_2$ global symmetry, with partition function 
\begin{equation}
(-1)^{\int_Y Sq^2[\beta_{d-1}]}
\end{equation}
Then $\tilde Z[X,\beta_{d-1}]$ defines a bosonic gapped boundary condition for the 
$(d+1)$-dimensional SPT phase, while $z[X,\eta,\beta_{d-1}]$ defines a fermionic 
gapped boundary condition. The original spin TQFT can be recovered by gauging 
the $(d-2)$-form $\ZZ_2$ global symmetry on a slab, with one of these boundary conditions at either end. 

In section \ref{sec:condensation} we will argue that this construction has a close relation to the notion of fermionic anyon condensation.
The generators of a non-anomalous $(d-2)$-form global symmetry $G$ are loop observables which can be thought 
as worldlines of bosonic quasi-particles which fuse accordingly to the group law of $G$.
Gauging the $(d-2)$-form symmetry is equivalent to proliferating these quasi-particles in correlation functions
and should correspond to the standard notion of anyon condensation, at least in three space-time dimensions. 

We will argue that the generators of a $(d-2)$-form $\ZZ_2$ global symmetry with the t'Hooft anomaly described above,
instead, behave as fermionic quasi-particles. The challenge to define a fermionic analogue of the standard
anyon condensation is mapped into the problem of gauging such anomalous symmetry. The kernel spin-TQFT $K_d$
offers a solution to the problem: given some TQFT with fermionic quasi-particles we want to condense, 
we can tensor it with $K_d$ to cancel the anomaly and gauge the $(d-2)$-form $\ZZ_2$ global symmetry.
Essentially, the kernel spin-TQFT $K_d$ uses the spin structure information to provide some extra signs 
which make the ``fermionic'' anyon condensation meaningful. 
This approach to fermionic anyon condensation appears to be closely related to work in progress by K. Walker~\cite{Walker}.  

State-sum constructions of spin-TQFTs in two dimensions have been recently discussed by other authors \cite{BarrettTavares,NovakRunkel}. 
It would be interesting to establish the precise connection between all these constructions. 

Finally, using the notion of fermionic anyon condensation we sketch a rough argument demonstrating 
how one could potentially ``simulate'' a generic fermionic lattice Hamiltonian in $2+1$ dimensions given a copy of $K_3$ 
and a sufficiently rich bosonic system. This argument supports the idea that $K_d$ may fully capture 
the non-local properties of a generic fermionic system.  

\subsection{Conclusions and future directions}
Although we have demonstrated the link between fermionic phases of matter and spin-TQFTs
only in a restricted set of examples, we believe that the relation will hold in greater generality. 
There are two natural ways one may try to extend our results
\begin{itemize}
\item It should be possible to adapt our analysis to the general mathematical framework of extended topological field theory. 
The analogue of a fermionic phase of matter should be an extended topological field theory 
such that the vector spaces attached to $(d-1)$-dimensional manifolds have Grassmann grading 
and the tensor products are twisted by the Koszul sign rule. The same sign combinatorics as in the 
state sum model should lead to an obstruction proportional to $[w_2]$. Thus we expect one could prove a theorem relating
``fermionic'' extended topological theories and extended spin-TQFTs. 
\item It should be possible to give a purely Hamiltonian version of our analysis.  The fermionic SPT phase $K_d$ associated to the 
$z[X,\eta,\beta_{d-1}]$ partition function should give us a recipe to build a one-dimensional Hilbert space 
from a given cocycle $\beta_{d-1}$, which could be described as a choice of sign on local patches of the space manifold. 
Hopefully, this recipe will capture the same sign ambiguities as one encounters in the construction of a general fermionic Hilbert space 
as a tensor product of local fermionic Hilbert spaces associated to local
patches of the space manifold. Ideally, this would show in full generality, without reference to TQFTs, 
that any fermionic phase of matter can be obtained by 
combining the fermionic SPT phase $K_d$ with some appropriate bosonic degrees of freedom and that fermionic phases of matter 
should generally require the existence of a spin structure on space. 
\end{itemize}
\section{Fermionic SPT phases and the Gu-Wen Grassmann integral}\label{sec:guwen}
\subsection{The Gu-Wen construction}

The standard discrete action for a bosonic SPT phase protected by some symmetry group $G$ 
is built from a $U(1)$-valued $d$- cocycle $\nu^b_d(g_0,\cdots, g_d)$ on BG, i.e. a function of 
$(d+1)$ $G$-valued variables, invariant under the action of $G$ on itself
\begin{equation}
\nu^b_d(g g_0,\cdots, g g_d) =  \nu^b_d(g_0,\cdots, g_d)
\end{equation}
and closed under the action of an appropriate differential $\delta$.

Concretely, given a triangulation of a $d$-dimensional manifold $X$ with a flat $G$ connection, one evaluates the partition function as a product over all 
$d$-dimensional simplices of $\nu_d^\pm$ evaluated on a local trivialization of the connection. The $G$-symmetry
of the cocycle makes the answer independent of the local trivialization and the cocycle condition 
$\delta \nu^b_d =1$ insures invariance under changes of triangulation. Essentially, the ratio between the partition functions for two triangulations 
which differ by an elementary move equals the partition function for the boundary of a $(d+1)$-simplex, which by definition is the same as $\delta \nu^b_d$.
More generally, two triangulations can be related by a sequence of moves which can be visualized as a triangulation of a cobordism $X \times [0,1]$ from the manifold $X$ to itself. 

The Gu-Wen construction of a discrete partition function for fermionic SPT phases \cite{GuWen} involves two basic pieces of input: 
a $U(1)$-valued $d$- cochain $\nu(g_0,\cdots, g_d)$ on BG and a $\ZZ_2$-valued $(d-1)$ cocycle $n_{d-1}(g_0, \cdots, g_{d-1})$,
such that 
\begin{equation}
\delta \nu_d = (-1)^{Sq^2[n_{d-1}]}
\end{equation}
In other words, $\nu_d$ satisfies the cocycle condition up to signs, which are determined from $n_{d-2}$ through the Steenrod
square operation. 

The Gu-Wen partition function can be decomposed into the product of three terms, 
which are not separately invariant under changes of triangulation. 
The first term, which we could denote as $Z_\nu$, or $Z_\nu[T]$ if we want to indicate the specific choice of triangulation $T$ of the space-time manifold $X$, 
is simply the product over all $d$-dimensional simplices of $\nu_d^{\pm 1}$,
just as for a bosonic SPT phase. Because $\nu_d$ is not a cocycle, the sign of this term will jump under re-triangulation 
by the Steenrod square of $n_{d-1}$ integrated over the cobordism from $X$ to itself:
\begin{equation}
Z_\nu[T] = Z_{\nu}[T'] \exp i \pi \int_{X \times [0,1]} Sq^2[n_{d-1}]
\end{equation}

The second term, which we could denote as $Z_\theta$ or $Z_\theta[T]$, contains the ``fermionic'' degrees of freedom. It is a sign, 
defined by a Grassmann integral whose structure is determined by $n_{d-1}$.
Schematically, one associate a pair of Grassmann odd variables to the two sides of each $(d-1)$-simplex such that 
$n_{d-1}$ is $1$. The integrand is built as a product over $d$-simplices of the Grassmann variables associated to 
that simplex. 
We do not expect to be able to write $Z_\theta$ as a standard bosonic action, i.e. the integral of the pull-back of 
some class on BG to the manifold: if we could do that, we would have reduced the system to a bosonic SPT phase. 
The Grassmann integral $Z_\theta$ only depends on the group variables through the image $\beta_{d-1}$
of the cochain $n_{d-1}$ computed on the faces of the triangulation, which is a standard $\ZZ_2$ cocycle. 
We will see later in Section \ref{sec:statesum} that the Grassmann integral $Z_\theta$
coincides with the function $\sigma_\Pi(X,\beta_{d-1})$ described in the 
introduction. 

The third term, which we could denote as $Z_m$ or $Z_m[T]$, is somewhat problematic: it is written in terms of a function 
$m_{d-2}(g_0, \cdots, g_{d-2})$ which is not $G$ invariant, but satisfies $\delta m_{d-2} = n_{d-1}$.
The expression for $Z_m$ involves a product of $(-1)^{m_{d-2}}$ evaluated over a certain subset $S$ of 
$(d-2)$-simplices in $T$ defined in \cite{GuWen} by some local rule:
\begin{equation}
Z_m[T] = \prod_{s \in S} (-1)^{m_{d-2}(s)} 
\end{equation}
We review the precise definition of $S$ for $d=2,3,4$ in 
section \ref{sec:S}. 

The product is invariant under re-definitions of $m_{d-2}$, but the lack of $G$-invariance of $m_{d-2}$ 
makes it problematic to define the model on a manifold with a non-trivial G-bundle. The wavefunctions built from this model have some 
$m_{d-2}$ dependence which is stripped off by hand by the authors of \cite{GuWen} in a non-canonical way, opening the possibility for 
subtle sign changes and ambiguities in the corresponding TQFT. Indeed, we will argue here that removing the $m_{d-2}$
dependence introduces naturally a dependence on a choice of spin structure on the manifold, so that the Gu-Wen fermionic SPT phases are a class of invertible spin-TQFTs.

In order to remove the spurious $m_{d-2}$ dependence, we can imagine replacing the collection $S$ of $(d-2)$-simplices used in $Z_m$
 with a collection of $(d-1)$ simplices $E$ such that $\partial E = S$, 
so that the product over $S$ of $(-1)^{m_{d-2}}$ in $Z_m$ can be reorganized to a product $Z_n^E$ over $E$ of $(-1)^{n_{d-1}}$:
\begin{equation}
Z_n^E[T] = \prod_{e \in E} (-1)^{n_{d-1}(e)}
\end{equation}
Clearly, this will be only possible if $S$ is exact, and then the result will depend on the choice of $E$:
any two different choices of $E$ differ by a cycle, which will have in general a non-zero 
pairing with $n_{d-1}$, and thus will give inequivalent partition functions.
For a solid torus, this integration by parts should 
mimick the way the authors of \cite{GuWen} strip off the $m_{d-2}$ dependence of wavefunctions. 

At this stage, we have shown that we can construct an improved, well-defined partition function if the homology class of 
$S$ vanishes, and that the improved partition function depends on a choice of trivialization $E$ of $S$. Although the definition of $S$ 
depends on the choice of triangulation, invariance of the partition function under changes of triangulation suggests that the 
homology class of $S$ should capture some intrinsic triangulation-independent property of the underlying space-time manifold 
$X$. 

We propose that the homology class of $S$ captures precisely the second Stiefel-Whitney class $[w_2]$, which vanishes 
on spin manifolds. We also propose that $S$ itself provides a canonical chain representative for $w_2$, 
so that the choice of $E$ actually encodes  a choice of a spin structure on the manifold.

The proposal implies that the partition function can only be  defined on a spin manifold, and depends on a choice of spin structure. 
The combination of signs $Z_\theta Z_n^E$ will coincide with the function $z[X,\eta,\beta_{d-1}]$ defined in the introduction.  
Conversely, if the cohomology class of $w_2$ is non-trivial, the partition sum will have an unavoidable dependence on re-definitions of 
$m_{d-2}$ which map to a non-trivial $(d-2)$ cohomology class on the manifold. 

We can offer two strong checks of our proposal: a general consistency check we describe momentarily, and a direct calculation 
for a special class of triangulations, which are obtained by refining a generic triangulation by a barycentric subdivision. 
These triangulations are endowed with a canonical representative chain for $w_2$, which turns out to coincide with $S$ for $d=2,3,4$.

The first consistency check follows from the observation that a change of $Z_m$ under a change of triangulation is a linear expression $\sum_{s\in V} m_{d-2}[s]$, 
a sum of $m$ evaluated on some collection $V$ of $(d-2)$-simplices of the cobordism $Y= X \times [0,1]$ from $X$ to itself.
The change of  $Z_m$ should only depend on $n_{d-1}$, as it is cancelled by the change in $Z_\nu$ and $Z_\theta$. 
It should thus be possible to write it as a linear expression $\sum_{s\in W} n_{d-1}[s]$, 
a sum of $n_{d-1}$ evaluated on some collection $W$ of $(d-1)$ simplices of $Y$ such that $\partial W = V$. 

The linear function defined by $W$ must be such that the sum 
\begin{equation}
\sum_{s\in W} n_{d-1}[s] + \int_Y Sq^2[n_{d-1}] 
\end{equation}
of the variations of $Z_m$ and $Z_\nu$ can be cancelled by the variation of $Z_\theta$.

There is a natural candidate for such a linear function: the integral of $w_2 \cup n_{d-1}$ over $Y$,
where $w_2$ is some specific cochain which represent the second Stiefel-Whitney class. A neat property of $w_2$
is that $Sq^2[x] + w_2 \cup x$ is exact for any $(d-1)$-cocycle $x$ and thus can be 
integrated by parts and plausibly cancelled by the variation of $Z_\theta$. If we write $n_{d-1} = \delta m_{d-2}$
and thus $w_2 \cup n_{d-1} = \delta (w_2 \cup m_{d-2})$, we find that $S$ must be a representative for $w_2$ on $X$. 

We can give a direct proof that $S$ is a representative for $w_2$ for a special class of triangulations $BT$, which are a barycentric subdivision 
of some rougher triangulation $T$. Each $d$-dimensional simplex $s$ of $T$ is subdivided in $(d+1)!$ simplices $s_\sigma$ 
whose vertices are the barycenters of subsets of the vertices of $s$. More precisely, given a permutation $\sigma$ of the 
vertices of $s$, the $i$-th vertex of $s_\sigma$ is the barycenter of the vertices $\sigma(0), \cdots, \sigma(i)$. 
Thus the vertices of $s_\sigma$ have a natural order, starting from a vertex of $s$ and ending with the barycenter 
of $s$. 

In the Gu-Wen construction, the triangulation is endowed with a branching structure, i.e. an orientation of the edges with no closed loops,
which in turn provides a order to the vertices of simplices: the vertex with all outgoing edges 
is number ``0'', the vertex with a single incoming edge is number ``1'', etcetera. 

If we have a barycentric triangulation, we can simply order the $1$-simplices from the barycenter of fewer vertices to the barycenter of more vertices. That gives a useful canonical choice of branching structure on $BT$, which induces the same order of the vertices of 
simplices as the natural one of the barycentric subdivision. This branching structure has the useful property that each vertex of $BT$ has the same position in the order of vertices in all simplices which include that vertex: vertices of $T$ are always at position ``0'' in the order, midpoint of segments in $T$ are always at position ``1'', etcetera.  

The barycentric subdivision $BT$ has an important property: the set of all $(d-2)$-simplices provides a chain which 
is a canonical representative for $w_2$. The collection $S$ of $(d-2)$ simplices used in \cite{GuWen} to define $Z_m$
is described as the sum over all $(d-2)$-simplices, plus a correction term given explicitly in dimensions $d=2,3,4$. 
We will now show by direct inspection that the extra correction term vanishes for a barycentric triangulation BT, so that $S$ is
precisely the canonical representative for $w_2$! 

\begin{figure}
\begin{center}
\includegraphics[width=10cm]{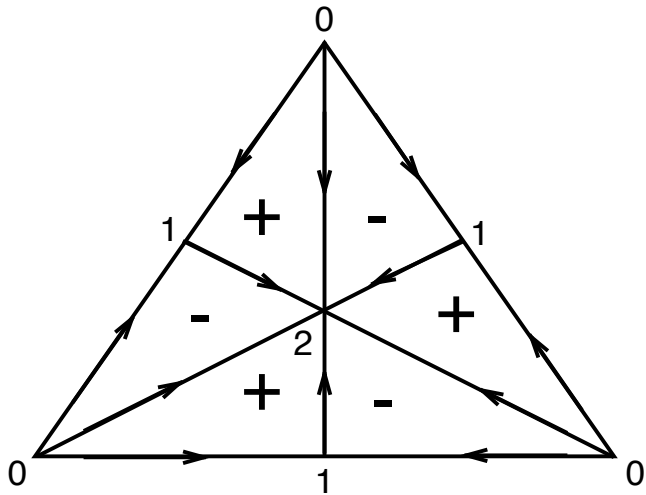}
\end{center}
\caption{The barycentric subdivision of a triangle}
\label{fig:bary}
\end{figure}

\subsection{Barycentric subdivisions in $d=2,3,4$} \label{sec:S}
\begin{figure}
\begin{center}
\includegraphics[width=10cm]{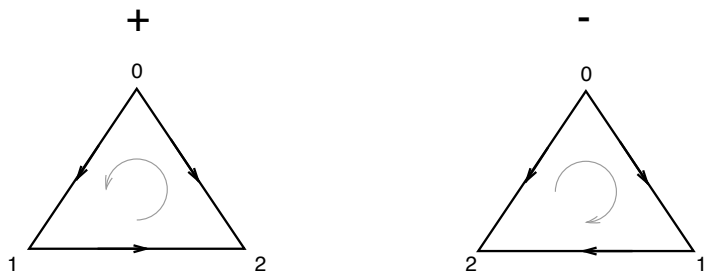}
\end{center}
\caption{The two possible orientations of a triangle and ordering of vertices induced by a branching structure on a 2d triangulation}
\label{fig:orient}
\end{figure}

For a general triangulation, the branching structure gives an order to the vertices of each triangle: the $n$-th vertex has n incoming edges.
The ordering of the vertices also gives an orientation of each simplex.
This orientation may or not agree with the canonical counterclockwise orientation of the simplex. 
If it does, we have a ``$+$'' simplex, if not we have a ``$-$'' simplex. See Figure \ref{fig:orient}.

In 2d, the partition function involves a factor of $(-1)^{m_0(g_v)}$ for each vertex $v$, and an extra factor of $(-1)^{m_0(g_v)}$
for each ``$-$'' triangle which has $v$ as the vertex number $1$. 

If we pick our triangulation to be a barycentric subdivision $BT$, then each ``1'' vertex belongs to two $-$ triangles, and 
we can disregard the second contribution. Thus $Z_m[BT]$ is the product over all vertices $v$ of $BT$ of $(-1)^{m_0(g_v)}$:
This is simply the pairing of $m_1$ with the canonical $0$-chain representative $S$ for $w_2$. 

Every 2d manifold admits a spin structure, so that $w_2$ is always exact. 
If we pick some 1-chain $E$ on $BT$ such that $\partial E = S$, we can replace $Z_m[BT]$
with the improved $Z_n^E[BT]$. We interpret the choice of $E$ as a choice of spin structure on the discretized surface. 

In 2d we can actually sketch a proof that $S$ is a chain representative for $w_2$ even for a generic triangulation $T$. 
In 2d, we can build representatives for $w_2$ by taking a vector field and picking the points where the vector field vanishes, 
counting how many times the vector field winds around the origin in a neighbourhood of each point, modulo $2$. That is the same as $1$ plus the 
number of times the vector field is tangent in the counterclockwise direction to a small circle around the point. If we pick the vector field $V$ in Figure 
\ref{fig:flow}, each vertex will contribute $1$ plus the number of times the vertex appears at position ``1'' in a $-$ triangle. That representative for $w_2$ 
coincides with $S$.

In 3d, the chain $S$ of edges which appear in $Z_m$ consists of all edges of the triangulation, together with the $(02)$ edge for all ``$+$'' tetrahedra and 
the $(13)$ edge for all ``$-$'' tetrahedra. 

For a barycentric triangulation $BT$, the $(02)$ edges join vertices of $T$ and barycenters of triangles in $T$. They are shared 
by two ``$+$'' triangles. The $(13)$ edges join midpoints of edges in $T$ and the barycenters of tetrahedra in $T$. They are shared 
by two ``$-$'' triangles. Thus the extra contributions to $S$ cancel out, and $S$ coincides with the sum of all edges, i.e. canonical chain representative for $w_2$. 

In 4d, the chain $S$ consists of all triangles in the triangulation, together with the $(013)$, $(134)$, $(123)$ triangles for a ``$+$'' 4-simplex, 
$(024)$ for a ``$-$'' 4-simplex. In order to show these edges are shared by an even number of tetrahedra of the corresponding 
orientation, we can observe that they are fixed points of two reflections. 

For example, a $(013)$ triangle 
has vertices which are barycenters of the first, the first $2$ and all vertices in a sequence of $4$ $(v_1,v_2,v_3,v_4)$. As it belongs to a face, 
it is shared by two tetrahedra in T. As is is invariant under exchange of $v_3$ and $v_4$, it is fixed by a second reflection.

Similarly, a $(134)$ triangle has vertices which are barycenters of the first $2$, $4$, $5$ vertices of a sequence of $5$ vertices. 
It is invariant under exchanging the first pair, or the second pair. A $(123)$ triangle has vertices which are barycenters of the first $2$, $3$, $4$ vertices of a sequence of $4$ vertices. It belongs to a face and is invariant under exchange of the first pair of vertices. Finally, a $(024)$ triangle has vertices which are barycenters of the first $1$, $3$, $5$ vertices of a sequence of $5$ vertices. It is invariant under exchange of the second and third vertices, and the fourth and fifth.

Thus all extra contributions to $S$ cancel out, and $S$ is given by the sum of all triangles, i.e. canonical chain representative for $w_2$.

In order to mimic in general dimension $d$ the 2d analysis for a general triangulation $T$, we would need to build a $w_2$ representative in terms of $d-1$ vector fields on $X$.
The representative would be concentrated at the locus where the vector fields fail to be linearly dependent, 
weighed by the number of times the vanishing linear combination of the vector fields winds around the other non-vanishing linear combinations in a neighbourhood of the locus. 
We expect it should be possible to define some canonical set of $d-1$ vector fields in each simplex, linearly dependent 
at $(d-2)$-simplices, which demonstrate the $S$ is a representative for $w_2$.   

\section{The Gu-Wen Grassmann integral as a quadratic refinement} \label{sec:quadratic}
\subsection{Quadratic property of the Gu-Wen Grassmann integral}

Next, we will look at $Z_\theta$. We aim to show that the combination 
\begin{equation}
Z_n^E[BT] Z_\theta[BT] \equiv z[X,\eta,\beta_{d-1}]
\end{equation}
is a quadratic function of the $\ZZ_2$ cocycle $\beta_{d-1}$, which refines the pairing on the space of $\ZZ_2$ $(d-1)$-dimensional cocycles
defined by the higher cup product $\cup_{d-2}$. 
 
The Grassmann integral only depends on the image $\beta_{d-1}(e)$ of
the cochain $n_{d-1}$ on the co-dimension one faces $e$ of the triangulation. 
Thus we can consider some generic $(d-1)$-cochain $\beta_{d-1}$, assigning a $\ZZ_2$ element to each edge of the triangulation,
and specialize it to the image of $n_{d-1}$ later on. 
For notational clarity, we will usually omit the subscript $(d-1)$ from $\beta$.

In the following formulae we will often refer to a cochain $A_p$ evaluated the simplex defined by vertices $a_0, \cdots a_p$ simply as $A_p(a_0 \cdots a_p)$.
We will also denote the Grassmann variables associated to an edge with vertices $a$ and $b$ as $\theta_{ab}$, $\bar \theta_{ab}$. 

We will start with a 2d example, and then proceed to higher dimensions. 
If $\beta(e)=1$ for an edge $e$, we assign to the edge Grassmann variables $\theta_e$ and $\bar \theta_e$: $\theta_e$ 
is associated to the side where the canonical orientation of the face agrees with the orientation of the edge, 
$\bar \theta$ on the side it disagrees. 

To each triangle $t$ we associate an even monomial $u(t)$, the product of the Grassmann variables attached to its edges, 
if present, in the order $12$, $01$, $02$ according to the ordering of the vertices for a $+$ triangle, opposite for a $-$ triangle.

Thus a $+$ triangle has monomial 
\begin{equation}
\theta_{12}^{\beta(12)} \theta_{01}^{\beta(01)} \bar \theta_{02}^{\beta(02)}
\end{equation}
and a $-$ triangle has 
\begin{equation}
\theta_{02}^{\beta(02)} \bar \theta_{01}^{\beta(01)} \bar \theta_{12}^{\beta(12)} 
\end{equation}

\begin{figure}
\begin{center}
\includegraphics[width=11cm]{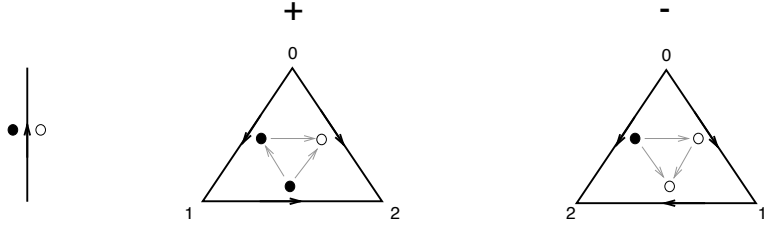}
\end{center}
\caption{
Left: to each edge $e$ such that $\beta(e)=1$ we assign a $\theta_e$ Grassmann variable (black dot) and a
$\bar \theta_e$ Grassmann variable (white dot). Each such edge contributes $d \theta_e d \bar \theta_e$ to the measure. 
Middle and right: as $\beta$ is a cochain, each triangle is associated to two Grassmann variables, which are ordered in the integrand 
according to the grey arrows.}
\label{fig:var}
\end{figure}

We are interested in the sign 
\begin{equation}
\sigma(\beta) = \int \prod_{e|\beta(e)=1} d \theta_e d \bar \theta_e \prod_t u(t)
\end{equation}

We will show directly from the definition that 
\begin{equation}
\sigma(\beta+ \beta') = \sigma(\beta) \sigma(\beta') (-1)^{\int_X \beta \cup \beta'} 
\end{equation}
i.e. $\sigma(\beta)$ is a quadratic function which refines the intersection pairing on $1$-cocycles. 

We can combine the two Grassmann integrals in $\sigma(\beta) \sigma(\beta')$ as
\begin{equation}
\sigma(\beta) \sigma(\beta')= \int \prod_{e|\beta(e)=1} d \theta_e d \bar \theta_e  \prod_{e|\beta'(e)=1} d \theta'_e d \bar \theta'_e \prod_t u(t)[\beta,\theta,\bar \theta] u(t)[\beta',\theta',\bar \theta'] 
\end{equation}
There is a subset of ``spurious'' Grassmann variables which is associated to edges for which $\beta(e)=\beta'(e)=1$. 
If we can systematically integrate them out, the remaining Grassmann variables will be associated to edges for which 
$\beta(e)+\beta'(e)=1$ and coincide with the Grassmann variables in the formula for $\sigma(\beta+ \beta')$.

Our strategy is to permute the variables until we form pairs $\theta_e \theta'_e$ and $\bar \theta_e \bar \theta'_e$ of spurious variables. 
These pairs are Grassmann even, and can be brought out of the integrand and eliminated against the 
$d \theta_e d \bar \theta_e d \theta'_e d \bar \theta'_e$ measure to give an overall $\prod_e (-1)^{\beta(e)\beta'(e)}$ sign.

The permutation of variables which brings the spurious pairs together is very simple: we interleave the 
variables in the product $u(t)[\beta,\theta,\bar \theta] u(t)[\beta',\theta',\bar \theta']$:
for a $+$ triangle we have 
\begin{align}
&\left[ \theta_{12}^{\beta(12)} \theta_{01}^{\beta(01)}  \bar \theta_{02}^{\beta(02)} \right]  \left[(\theta_{12}')^{\beta'(12)} (\theta_{01}')^{\beta'(01)} (\bar \theta'_{02})^{\beta'(02)} \right] =\cr &
= (-1)^{\beta(01)\beta'(12)+ \beta(02)\beta'(12) + \beta(02)\beta'(01)} \cdot \cr & \cdot \left[ \theta_{12}^{\beta(12)}(\theta_{12}')^{\beta'(12)} \right] \left[ \theta_{01}^{\beta(01)}(\theta_{01}')^{\beta'(01)} \right] \left[ \bar \theta_{02}^{\beta(02)} (\bar \theta'_{02})^{\beta'(02)} \right]
\end{align}
and for a $-$ triangle we have 
\begin{align}
&\left[ \theta_{02}^{\beta(02)} \bar \theta_{01}^{\beta(01)} \bar \theta_{12}^{\beta(12)}  \right]  \left[(\theta_{02}')^{\beta'(02)} (\bar \theta_{01}')^{\beta'(01)} (\bar \theta'_{12})^{\beta'(12)} \right] =\cr &
= (-1)^{\beta(01)\beta'(02)+ \beta(12)\beta'(02) + \beta(12)\beta'(01)} \cdot \cr & \cdot \left[ \theta_{02}^{\beta(02)} (\theta'_{02})^{\beta'(02)} \right] \left[ \bar \theta_{01}^{\beta(01)}(\bar \theta_{01}')^{\beta'(01)} \right]\left[ \bar \theta_{12}^{\beta(12)}(\bar \theta_{12}')^{\beta'(12)} \right]
\end{align}
Thus we have grouped together pairs of spurious variables, if present, and left the non-spurious variables in the correct order to be identified with $u(t)[\beta+ \beta',\cdots]$.

It is useful to re-distribute the overall $\prod_e (-1)^{\beta(e)\beta'(e)}$ sign, by associating each factor of $(-1)^{\beta(e)\beta'(e)}$ with the triangle to the right of the oriented edge $e$. 
Thus each $+$ triangle, which only sits to the right of its $02$ edge, is associated to an overall factor 
\begin{equation}
(-1)^{\beta(01)\beta'(12)+ \beta(02)\beta'(12) + \beta(02)\beta'(01) + \beta(02)\beta'(02)}  = (-1)^{\beta(01)\beta'(12)}  
\end{equation}
and to each $-$ triangle, which only sits to the right of its $01$ and $12$ edges, is associated to an overall factor 
\begin{equation}
(-1)^{\beta(01)\beta'(02)+ \beta(12)\beta'(02) + \beta(12)\beta'(01) + \beta(01)\beta'(01)+ \beta(12)\beta'(12)}  = (-1)^{\beta(01)\beta'(12)}  
\end{equation}

Thus the reorganization of the Grassmann integral $\sigma(\beta) \sigma(\beta')$ into the Grassmann integral $\sigma(\beta+ \beta')$ produces a sign of 
$(-1)^{\beta(e_{01})\beta'(e_{12})} $ for each triangle. This is nothing else but the cup product of $\beta$ and $\beta'$!
Thus we have the desired quadratic property
\begin{equation}
\sigma(\beta+ \beta') = \sigma(\beta) \sigma(\beta') (-1)^{\int_X \beta \cup \beta'} 
\end{equation}

We can readily extend the above argument to $Z_\theta$ in any dimension $d$. 
For general $d$, the Grassmann integral involves again pairs of variables associated to the two sides of each 
oriented $(d-1)$-simplex $e$ such that $\beta(e)=1$. 

The integrand is a product of monomials $u[s]$ made out of all Grassmann variables associated to each $d$-simplex $s$. 
The order of the variables in the monomial is determined by a specific rule. We will describe the rule later, for now we only need to 
know that it gives a canonical order to the faces of each simplex. 

If we denote as $\sigma(\beta)$ again the result of the Grassmann integral, the same interleaving operation 
as in the 2d case gives us immediately a proof that $\sigma(\beta)$ is quadratic: 
\begin{equation}
\sigma(\beta+ \beta') = \sigma(\beta) \sigma(\beta') \prod_s \epsilon[s,\beta, \beta'] 
\end{equation}
with 
\begin{equation}
\epsilon[s,\beta, \beta'] = (-1)^{\sum_{e,e' \in s}^{e>e'} \beta(e) \beta'(e') + \sum_{e\in s}^{e>0} \beta(e) \beta(e')}
\end{equation}
with $e>e'$ in the order determined by $u[s]$ and $e>0$ if $u[s,\beta]$ includes a $\bar \theta_e$ variable.
We aim to identify the $d$-cochain in the exponent with the higher cup product $\beta \cup_{d-2} \beta'$.

The order of faces induced by $u[t]$ for a $+$ simplex places first the faces which omit the even vertices 
(first the one omitting $0$, then the one omitting $2$, etc.), which are labelled by $\theta_e$ variables, and then 
the faces which omit the odd vertices (first the one omitting $1$, then the one omitting $3$, etc.), which are labelled by $\bar \theta_e$ variables.
The order for a $-$ simplex is the opposite (and the role of $\theta_e$ and $\bar \theta_e$). 

We will now compute $\epsilon[s,\beta, \beta']$ in $d=3$ and $d=4$ for this choice of order and verify it is given by the standard higher cup products $\beta \cup_1 \beta'$
and $\beta \cup_2 \beta'$ respectively. A similar analysis in higher dimension should be straightforward. 

Notice that if we were to pick a different choice of order $\Pi$ for the factors in the measure, $\sigma_\Pi(\beta)$ would differ from the standard $\sigma(\beta)$ by a local linear term, 
which does not affect the quadratic refinement property. If we pick a different choice of order $\Pi$ for the factors in $u[t]$, $\sigma_\Pi(\beta)$ will differ from the standard $\sigma(\beta)$
by a local quadratic term, some $(\beta,\beta)$ pairing defined by a sum over simplices of the product of fermion numbers of the variables which 
have been permuted in $u[t]$ to get to the new order. That would change the quadratic refinement from $\beta \cup_{d-2} \beta'$
to some other 
\begin{equation}
\beta \tilde \cup_{d-2} \beta' = \beta \cup_{d-2} \beta' + (\beta,\beta')+ (\beta',\beta)
\end{equation}
This is just a different choice of definition for the higher cup product: the basic relation in equation \ref{eq:highercup}
remains valid with the re-definition 
\begin{equation}
\beta \tilde \cup_{d-3} \beta' = \beta \cup_{d-2} \beta' + \delta (\beta,\beta')+(\delta \beta,\beta') + (\beta, \delta \beta')
\end{equation}
and all other products unchanged. 

\subsection{Quadratic refinement in $3d$}
At a $+$ simplex, we have an order of faces $(123)$, $(013)$, $(023)$, $(012)$ and thus 
the signs computed from the quadratic pairing 
\begin{align}
&\beta(012) \beta'(123) + \beta(012) \beta'(013) + \beta(012) \beta'(023) + \beta(023) \beta'(123) +\cr & +\beta(023) \beta'(013)+  \beta(013) \beta'(123) + \beta(023) \beta'(023) + \beta(012) \beta'(012) = \cr
&\beta(023) \beta'(012) + \beta(013) \beta'(123) 
\end{align}
At a $-$ simplex, we have an order of faces $(012)$, $(023)$, $(013)$, $(123)$,  and thus 
the signs computed from the quadratic pairing 
\begin{align}
&\beta(123) \beta'(012) + \beta(123) \beta'(023) + \beta(123) \beta'(013) + \beta(013) \beta'(012) + \cr & + \beta(013) \beta'(023)+ \beta(023) \beta'(012) + \beta(013) \beta'(013) + \beta(123) \beta'(123) = \cr
&\beta(023) \beta'(012) + \beta(013) \beta'(123) 
\end{align}
Again, the dependence on the type of triangle drops out. 
We have a $d$-cochain $(\beta, \beta')_3$ given by the pairing $\beta(023) \beta'(012) + \beta(013) \beta'(123)$.
This is precisely the definition of $\beta \cup_1 \beta'$. Thus we claim that 
in 3d we have 
\begin{equation}
\sigma(\beta+ \beta') = \sigma(\beta) \sigma(\beta') (-1)^{\int_X \beta \cup_1 \beta'} 
\end{equation}

\subsection{Quadratic refinement in $4d$}
At a $+$ simplex, we have an order of faces $(1234)$, $(0134)$, $(0123)$,$(0234)$, $(0124)$ and thus 
the signs computed from the quadratic pairing 
\begin{align}
&\beta(0124)\left[ \beta'(1234)+\beta(0134)+\beta(0123)+ \beta(0234) \right] + \cr &+\beta(0234)\left[ \beta'(1234)+\beta'(0134)+\beta'(0123)\right]  +\cr &+\beta(0123)\left[ \beta'(1234)+\beta'(0134)\right] + \beta(0134)\beta'(1234) +\cr &+ \beta(0124)\beta'(0124) + \beta(0234)\beta'(0234) =\cr &\beta(0234)\beta'(0124) +\beta(0123)\beta'(1234)+\cr &+\beta(0123)\beta'(0134) + \beta(0134)\beta'(1234) 
\end{align}
At a $-$ simplex, we find the same. 
Thus have a $d$-cochain 
\begin{align}(\beta, \beta')_4=&\beta(0234) \beta'(0124) + \beta(0123) \beta'(1234)+\cr &+\beta(0123) \beta'(0134) + \beta(0134) \beta'(1234) \end{align}
which is the standard expression for $\beta \cup_2 \beta'$.

\section{The t'Hooft anomaly}\label{sec:anomaly}
\subsection{Gauge variation of the Gu-Wen Grassmann integral }

Next, we can look at the variation of the Grassmann integral under exact changes in the cocycle $\beta_{d-1}$.
The calculation is greatly simplified by the quadratic refinement property. 
\begin{equation}
\sigma(\beta+ \delta \lambda) = \sigma(\beta) \sigma(\delta \lambda) (-1)^{\int_X \beta \cup_{d-2} \delta \lambda}  = \sigma(\beta) \sigma(\delta \lambda) (-1)^{\int_X \beta \cup_{d-3} \lambda + \lambda \cup_{d-3} \beta}
\end{equation}
where we used the basic property of the higher cup product, i.e. eqn. \ref{eq:highercup}: the violation of the Leibniz rule for $\cup_a$ equals the symmetrization of $\cup_{a-1}$. 

We can specialize the quadratic refinement property to 
\begin{equation}
\sigma(\delta \lambda+ \delta \lambda')=
\sigma(\delta \lambda) \sigma(\delta \lambda') (-1)^{\int_X \lambda \cup_{d-3} \delta  \lambda' + \lambda' \cup_{d-3}  \delta \lambda + \lambda \cup_{d-4} \lambda'+ \lambda' \cup_{d-4} \lambda}
\end{equation}
which can be solved up to a linear ambiguity: 
\begin{equation} \label{eq:anomalytilde}
\sigma(\delta \lambda)= (-1)^{\sum_{s \in \tilde S} \lambda(s) + \int_X \lambda \cup_{d-3} \delta  \lambda + \lambda \cup_{d-4} \lambda}
\end{equation}

In order to fix the ambiguity, it is useful to compute directly $\sigma(\delta \lambda)$ for the simplest possible case, where $\lambda$ is non-zero only on a single 
$(d-2)$-simplex, so that $\delta \lambda$ equals $1$ on all the $(d-1)$-simplices whose boundary include the 
selected $(d-2)$-simplex. Thus if we go around the $(d-2)$-simplex, we will encounter a sequence of Grassmann variables 
$\vartheta_i$, with measure factors $\pm d \vartheta_{2i} d \vartheta_{2i+1}$ from the $(d-1)$-simplices and 
$u(t)$ factors $\pm \vartheta_{2i+2} \vartheta_{2i+1}$ from the $d$-simplices inserted around the $(d-2)$-simplex.

If all the signs above were $+$, the overall Grassmann integral would give a factor of $-1$. 
In general, the signs in the measure factors are determined by the orientation of the $(d-1)$-simplices 
and the signs in the integrand factors $u[t]$ are determined by the relative order of the two $(d-1)$-simplices Grassmann variables
in $u[t]$. 

Consider for simplicity a barycentric subdivision. We have several types of $(d-2)$ simplices, which can be labellet by two integers from $0$ to $d$:
they are simplices whose vertices do not include the barycenters of $a$ vertices or of $b$ vertices. They are associated to an alternating sequence of 
$(d-1)$ simplices which omit either barycenters of $a$ vertices or of $b$ vertices. As these simplices are also of alternating type $+$ or $-$, 
it is easy to see that we can pick a direction around the $(d-2)$-simplex such that the integrand factors will have all $+$ signs. 
Working through a few examples, it is easy to convince oneself that the signs in the measure factors multiply to $1$. 
Thus if $\lambda_s$ is non-zero only on a single $(d-2)$-simplex $s$ in $BT$,
we can write 
\begin{equation}
\sigma(\delta \lambda_s) =-1
\end{equation} 

As the quadratic part of equation \ref{eq:anomalytilde} vanishes for $\lambda_s$, we find that $\tilde S$ for a barycentric subdivision $BT$ coincides again with the 
canonical representative for $w_2$. We expect $\tilde S$ to essentially coincide with $S$ and the canonical chain representative for $w_2$ for a general triangulation as well. 
In other words, we expect that 
\begin{equation}
z[X,\eta,\delta \lambda_s] =1
\end{equation}

Thus we can write 
\begin{equation} \label{eq:anomaly}
\sigma(\delta \lambda)= (-1)^{ \int_X \lambda \cup_{d-3} \delta  \lambda + \lambda \cup_{d-4} \lambda + w_2 \cup \lambda}
\end{equation}
and thus:
 \begin{equation}
\sigma(\beta+ \delta \lambda) =\sigma(\beta) (-1)^{\int_X \beta \cup_{d-3} \lambda + \lambda \cup_{d-3} \beta + \lambda \cup_{d-3} \delta  \lambda + \lambda \cup_{d-4} \lambda + w_2 \cup \lambda}
\end{equation}

We can also write the 't Hooft anomaly of $z[X,\eta,\beta_{d-1}]$ under gauge transformations of $\beta_{d-1}$:
 \begin{equation}
z[X,\eta,\beta_{d-1}+ \delta \lambda_{d-2}] =z[X,\eta,\beta_{d-1}] (-1)^{\int_X \beta \cup_{d-3} \lambda + \lambda \cup_{d-3} \beta + \lambda \cup_{d-3} \delta  \lambda + \lambda \cup_{d-4} \lambda}
\end{equation}

\subsection{A WZW-like expression for a quadratic refinement}
For $d\geq 3$ one can construct a WZW-like expression for the quadratic function of the cocycle $\beta$ as follows. 
Let us assume that $X$ is a boundary of some compact oriented $(d+1)$-manifold $Y$. 
This is automatic if $d\leq 3$, since the oriented bordism group $\Omega_d^{SO}(pt)$ vanishes for $d=2,3$ \cite{Stong}, but in general it is a nontrivial constraint on $X$. 
If we are given a $(d-1)$-cocycle $\beta\in Z^{d-1}(X,\ZZ_2)$ on $X$, one can always choose $Y$ so that $\beta$ extends to a $(d-1)$-cocycle on $Y$. 
To see this, we regard $\beta$ as map $\beta: X\ra K(\ZZ_2,d-1)$ where $K(\ZZ_2,d-1)$ is an Eilenberg-MacLane space; then the statement we need is that the reduced oriented bordism group 
$\tilde\Omega_d^{SO}(K(\ZZ_2,d-1))$ vanishes. This follows from the Atiyah-Hirzebruch spectral sequence for (unreduced) bordism and the vanishing of the reduced homology of $K(\ZZ_2,d-1)$ in degree less than $d-1$. 
Moreover,  if $d\geq 3$, the reduced oriented $d$-dimensional bordism of a product of several copies of $K(\ZZ_2,d-1)$ also vanishes, for the same reason.\footnote{This fails for $d=2$, because of the ${\rm Tor}$ terms in the K\"unneth formula. In fact, it is easy to give an example of a closed oriented  2-manifold $X$ and a pair of classes $\alpha_1,\alpha_2\in H^1(X,\ZZ_2)$ such that there is no compact oriented 3-manifold $Y$ such that $X=\partial Y$ and both $\alpha_1,\alpha_2$ arise from restriction of  cohomology classes on $Y$. For example, one can take $X$ to be a 2-torus, with $\alpha_1$ and $\alpha_2$ being the generators of $H^1(X,\ZZ_2)$.}
This implies that one can choose $Y$ to be independent of $\beta$. Such a $Y$ is not unique, of course.

Consider now the following WZW-like expression:
\begin{equation}
\tilde \sigma(\beta) = (-1)^{\int_Y \beta \cup_{d-3} \beta + w_2 \cup \beta}
\end{equation}
It is independent of the actual choice of $Y$, as $\beta \cup_{d-3} \beta + w_2 \cup \beta$ is known to be exact if $\beta$ is a cocycle.

This expression is a quadratic refinement of $\cup_{d-2}$. Indeed, if $\beta$ and $\beta'$ are $(d-1)$-cocycles on $X$, we get
\begin{equation}
\tilde \sigma(\beta+ \beta') = \tilde \sigma(\beta) \tilde \sigma(\beta') (-1)^{\int_X \beta \cup_{d-2} \beta'}
\end{equation}

It also transforms in the same way as $\sigma$ under 1-form $\ZZ_2$ gauge symmetry:
\begin{align}
\tilde \sigma(\delta \lambda) &= (-1)^{\int_Y \delta \lambda \cup_{d-3} \delta \lambda + w_2 \cup \delta \lambda} = \cr
&= (-1)^{\int_Y \lambda \cup_{d-4} \delta \lambda + \delta \lambda \cup_{d-4} \lambda + \int_X \lambda \cup_{d-3} \delta \lambda + w_2 \cup \lambda}  \cr
& =(-1)^{ \int_X \lambda \cup_{d-3} \delta \lambda + \lambda \cup_{d-4} \lambda + w_2 \cup \lambda}.
\end{align}
This means that  $\sigma$ and $\tilde \sigma$ can only differ by a linear and gauge-invariant function of $\beta$.

Thus for $d\geq 3$ for all practical purposes we can write
 \begin{equation}
z[X,\eta,\beta_{d-1}] =(-1)^{\int_X \eta \cup \beta_{d-1} + \int_Y \beta_{d-1} \cup_{d-3} \beta_{d-1} + w_2 \cup \beta_{d-1}}
\end{equation}

\section{Fermionic SPT phases and spin cobordism}

It was proposed in \cite{Kap:cobord} that fermionic Short Range Entangled phases in $d$ space-time dimensions with symmetry $G$ and vanishing thermal Hall conductivity are classified by the Pontryagin dual of the torsion part of $\Omega_d^{Spin}(BG)$. Here $BG\simeq K(G,1)$ is the classifying space of $G$. Some checks of this were performed in \cite{FSPT}. We can now compare with the Gu-Wen supercohomology proposal in low dimensions. 

Let us begin with $d=3$. In this dimension there are no nontrivial fermionic SRE phases in the absence of symmetry, so in the presence of symmetry there is no distinction between SRE and SPT phases. From the mathematical viewpoint, one has $\Omega_3^{Spin}(pt)=0$, and thus $\Omega_3^{Spin}(BG)$ coincides with the reduced bordism group $\tilde\Omega_3^{Spin}(BG)$. The partition function of the  model corrected by the spin-structure dependent term is
\begin{equation}
Z(X,A,\eta)=\exp\left(2\pi i\int_X A^*\nu_3\right) z(X,\eta,A^*\beta_2),
\end{equation}
where $\nu_3\in C^3(BG,\RR/\ZZ),$ $\beta_2\in Z^2(BG,\ZZ_2)$, and $A$ is a gauge field on $X$ regarded as a map $A:X\ra BG$. The cochains $\mu_3$ and $\beta_2$ satisfy
\begin{equation}
\delta\nu_3=\frac12 \beta_2\cup\beta_2.
\end{equation}
It is easy to see that this expression defines an element of the Pontryagin dual of $\Omega_3^{Spin}(BG)$. Indeed, it is clear that $Z(A,\eta)$ is multiplicative under disjoint union. Now, suppose there exists a compact spin 4-manifold $Y$ with boundary $(X,\eta)$ such that $A$ extends to a map $A_Y: Y\ra BG$. Then
\begin{equation}\label{nuthree}
\exp(2\pi i\int_X A^*\nu_3)=(-1)^{\int_Y A_Y^*\beta_2\cup A_Y^*\beta_2}.
\end{equation}
On the other hand, the WZW-like expression for $z(X,\eta,\beta_2)$ becomes
\begin{equation}
(-1)^{\int_X \eta\cup A^*\beta_2+\int_Y w_2\cup A_Y^*\beta_2+\int_Y A_Y^*\beta_2\cup A_Y^*\beta_2},
\end{equation}
which is clearly the same as (\ref{nuthree}). Since $Z(X,A,\eta)$ becomes $1$ when evaluated on trivial bordism classes, it defines a homomorphism from $\Omega_3^{Spin}(BG)$ to $U(1)$. 

Not all spin cobordism classes can be so obtained. For example, for $G=\ZZ_2$ it is known that $\Omega_3^{Spin}(B\ZZ_2)\simeq\ZZ_8$ \cite{FSPT}, while the Gu-Wen construction only gives phases labeled by $\ZZ_4$. From the physics side, it is also known that 3d fermionic SPT phases with $\ZZ_2$ symmetry are classified by $\ZZ_8$ \cite{LevinGu}.

For $d>3$ there may exist nontrivial fermionic SRE phases even in the absence of any symmetry. According to \cite{Kap:cobord}, they exist whenever $\Omega_d^{Spin}(pt)$ has torsion. If we want to focus on fermionic SPT phases, we can restrict to $d$-manifolds $X$ which define a trivial class in $\Omega_d^{Spin}(pt)$. Then the same argument shows that if $A$ extends to a compact spin $(d+1)$-manifold $Y$ such that $\partial Y =X$, then the partition function of the Gu-Wen model is $1$. Therefore, each Gu-Wen supercohomology class defines a homomorphism from the reduced bordism group $\tilde\Omega_d^{Spin}(BG)$ to $U(1)$. Again, in general we do not expect that all such homomorphisms can be obtained from the Gu-Wen construction.

In $d=2$ the arguments are a bit different, since there is no WZW-like expression for $z(X,\eta,\beta_1)$. The proposal of \cite{Kap:cobord} is that fermionic SPT phases are classified by the Pontryagin dual of the reduced bordism group $\tilde\Omega_2^{Spin}(BG)$. There are also fermionic SRE phases in the absence of any symmetry which are classified by the dual of $\Omega_2^{Spin}(pt)=\ZZ_2$. The Gu-Wen construction describes only the former. To describe the correspondence, recall that a spin structure on an oriented 2-manifold can be identified with a quadratic refinement of the intersection form on $H_1(X,\ZZ_2)\simeq H^1(X,\ZZ_2)$ \cite{Atiyah,Johnson}. This quadratic refinement is nothing but $z(X,\eta,\beta_1)$, see Appendix A for a detailed discussion. Moreover, it follows from the results of \cite{Atiyah} that the value of $z(X,\eta,\beta_1)$ depends only on the bordism class of $(X,\eta,\beta_1)$ in $\Omega_2^{Spin}(B\ZZ_2)$. Thus the Gu-Wen construction defines a map from $H^1(BG,\ZZ_2)$ to spin cobordism of $BG$. This map is not a homomorphism, because $z(X,\eta,\beta_1)$ is not linear but quadratic in $\beta_1$. But we should remember that every element $\nu_2\in H^2(BG,\RR/\ZZ)$ also gives us an element in the spin cobordism of $BG$. Thus Gu-Wen SPT phases are described by pairs $(\nu_2,\beta_1)\in H^2(BG,\RR/\ZZ)\times H^1(BG,\ZZ_2)$. The group structure is a nontrivial extension of $H^1(BG,\ZZ_2)$ by $H^2(BG,\RR/\ZZ)$:
\begin{equation}\label{extension}
(\nu_2,\beta_1)+(\nu_2',\beta_1')=(\nu_2+\nu_2'+\frac12 \beta_1\cup\beta_1',\beta_1+\beta_1').
\end{equation}
It follows from the Atiyah-Hirzebruch spectral sequence that this extension is isomorphic to the dual of the reduced bordism $\tilde\Omega_2^{Spin}(BG)$, in agreement with the proposal of \cite{Kap:cobord}. 

The nontrivial $d=2$ fermionic SRE phase without any symmetry is realized by the Kitaev spin chain. The corresponding spin-TQFT has the Arf invariant as its partition function (the unique bordism invariant of spin structures in $d=2$). We recall that the Arf invariant of $(X,\eta)$ is essentially the average of the $z(X,\eta,\beta_1)$ over all 
$\beta_1\in H^1(X,\ZZ_2) $ \cite{Johnson}.  This spin-TQFT can also be constructed using the Gu-Wen Grassmann integral, see section \ref{sec:tv}.

\section{Constructing spin-TQFTs in low dimensions}\label{sec:tv}

\subsection{State-sum constructions of 2d spin-TQFTs}

An oriented 2d TQFT can be defined axiomatically as a functor from a geometric category $\Cob_2$ to  the category of vector spaces. The category $\Cob_2$ has  closed $1$-manifolds as objects and oriented bordisms between them as morphisms. It is well known that there is a 1-1 correspondence between oriented 2d TQFTs and commutative Frobenius algebras. The commutative Frobenius algebra corresponding to a given 2d TQFT encodes 2-point and 3-point functions on a sphere, and the rest of the correlators can be reconstructed from it. 

An alternative approach to constructing 2d TQFTs is provided by state-sum models \cite{latticeTQFT,BP}. This approach is more natural from the statistical mechanics viewpoint and gives a manifestly local recipe for computing the TQFT partition function and correlators for a triangulated closed oriented $d$-manifold $X$. One starts with a not necessarily commutative semi-simple Frobenius algebra $A$ and defines the partition function as follows. Fix a basis $e_i,$ $i\in I$, in $A$ and denote by $C^i_{jk}$ the structure constants of $A$ in this basis. Let
$$
g_{ij}=C^l_{ik} C^k_{jl}.
$$
The matrix $g_{ij}$ is non-degenerate if $A$ is semi-simple. Let $g^{ij}$ be its inverse. It is easy to see that
$$
C_{ijk}=g_{il} C^l_{jk}
$$
is cyclically symmetric. A coloring of a 2-simplex $f$ of $X$ is an assignment of an element of $I$ to each boundary 1-simplex of $f$. A coloring of a triangulation is a coloring of each 2-simplex. If each 2-simplex is colored, each 1-simplex has two colors. The partition function of a triangulated manifold $X$ is a sum over all colorings of the triangulation, with the weight of each coloring defined as product of weights of 1-simplices and 2-simplices. The weight of a 1-simplex colored by $i,j\in I$ is $g^{ij}$.
The weight of a 2-simplex whose three edges are colored by $i,j,k$ is $C_{ijk}$. Here we use the cycling ordering of the edges arising from the orientation of $X$. It is easy to show that the partition function thus defined is independent of the choice of a triangulation \cite{latticeTQFT,BP}.

The state-sum construction is somewhat redundant, as the partition function and the correlators depend only on the center $Z(A)$, which is a semi-simple commutative Frobenius algebra. The algebra $A$ can be interpreted as the algebra of boundary operators for a particular boundary condition for the TQFT based on $Z(A)$. On the other hand, the state-sum construction is very explicit and can be easily extended to manifolds with boundaries. 

Note that the state-sum construction always gives rise to a semi-simple $Z(A)$ and therefore does not produce the most general oriented 2d TQFT. For applications to condensed matter physics, this is not a serious drawback, since unitary TQFTs are automatically semi-simple.  

It is easy to modify the state-sum construction to produce 2d spin-TQFTs. One starts with a $\ZZ_2$-graded semi-simple algebra $A$. Let $e_i,$ $i\in I$, be its basis.
Each basis vector is assumed to have a well-defined grading $\beta(i)\in \ZZ_2$. As in the bosonic case, we color each edge of each 2-simplex with an element of $I$, so that each 1-simplex is colored with a pair of elements of $I$. The weights assigned to 1-simplices and 2-simplices will be proportional to $g^{ij}$ and $C_{ijk}$. Since the algebra $A$ is $\ZZ_2$-graded, the matrix $g^{ij}$ vanishes if $\beta(i)\neq\beta(j)$. Thus we may assume that if a 1-simplex is colored by $i,j$, then $\beta(i)=\beta(j)$. Thus each allowed coloring defines a $\ZZ_2$-valued 1-cochain $\beta$. Since $C_{ijk}$ vanishes unless $\beta(i)+\beta(j)+\beta(k)=0$, this 1-cochain is a cocycle. 

In order to account for the Grassmann nature of elements of $A$, we correct the naive sum over the colour of edges by including an overall sign.
This is the sign which arises from identifying $C_{ijk}$ with an element of $A^* \otimes A^* \otimes A^*$ and $g^{ij}$ with an element of $A \otimes A$:
we pick some order of the factors in each individual weight and then re-order them to bring together the pairs of $A$ and $A^*$ spaces we want to contract. 

The combinatorics of the Koszul signs is the same as in the Gu-Wen construction. The Grassman variables $\theta_e$, $\bar \theta_e$ act as placeholders for the Grassman-odd 
generators of $A^*$ in the weights of 2-simplices and the $d \theta_e$, $d \bar \theta_e$ as placeholders for the Grassmann-odd 
generators of $A$ in the weights of 1-simplices. The contraction between generators of $A$ and $A^*$ is mimicked by the Grassman integration. 
The result is just $\sigma(\beta)$. 

Thus the naive weight of each allowed coloring will be the product of $g^{ij}$ over 1-simplices, $C_{ijk}$ over 2-simplices and $\sigma(\beta)$. 
We can rewrite the sum over colorings  as a sum over colorings producing a particular 1-cocycle $\beta$ followed by a sum over $\beta$. Let us denote by $Z[\beta]$ the result of the first summation. 
The discussion in section 2 implies that $Z[\beta]$ is independent of the triangulation up to a sign. To make it completely independent, we need to choose a trivialization $\eta$ of $w_2$ and
multiply $Z[\beta]$ by a correction factor
$$
(-1)^{\int_X \eta\cup \beta}.
$$
i.e. we take the correct weight to be the product of $g^{ij}$ over 1-simplices, $C_{ijk}$ over 2-simplices and $z(X,\eta,\beta)$. 
Then the partition partition function depends on the spin structure $\eta$ on $X$, but not on a particular triangulation.
Notice that in 2d $z(X,\eta,\beta)$ coincides with the well-known quadratic refinement $(-1)^{q_\eta(\beta)}$ of the intersection form, 
evaluated on $\beta$ (see Appendix  A).

The simplest example of a 2d  spin-TQFT is obtained if we take $A$ to be the Clifford algebra ${\rm Cl}(1)$. 
It is generated by $1$ and an odd variable $\eta$ satisfying $\eta^2=1$. 
The $2\times 2$ matrix $g_{ij}$, $i\in \ZZ_2$, is given by $g_{ij}=2\delta_{ij}$, while $C_{ijk}$ is equal to either $2$ or $0$ depending on whether the sum of the indices is $0$ or $1$ modulo $2$.  
In this case the coloring is completely determined by the 1-cochain $\beta$, and the weight is simply $z(X,\eta,\beta) = (-1)^{q_\eta(\beta)}$. 
More precisely, if we denote by $E$ and $F$ the number of 1-simplices and 2-simplices respectively, the partition function reduces to
$$
2^{F-E} \sum_\beta (-1)^{q_\eta(\beta)}=2^{1-g(X)} \Arf(X,\eta),
$$
where $g(X)$ is the genus of $X$, and $\Arf(X,\eta)\in \{\pm 1\}$ is the  Arf invariant of the spin structure $\eta$ \cite{Atiyah,Johnson,Quadratic}:
$$
\Arf(X,\eta)=2^{-g(X)} \sum_{[\beta]\in H^1(X,\ZZ_2)} (-1)^{q_\eta(\beta)}.
$$
The partition function of the state-sum is not $\pm 1$, but it differs from it by an exponential of an integral of a local counter-term (the Euler density), and thus can be made
$\pm 1$ by a local redefinition. After such a redefinition, we get an invertible 2d spin-TQFT which describes the basic fermionic SRE phase in two space-time dimensions (the Mayorana spin chain). This spin-TQFT has also been briefly discussed in \cite{MooreSegal}.

\subsection{State-sum construction of 3d spin-TQFTs} \label{sec:statesum}

In 3d, the state-sum construction of bosonic TQFTs is known as the Turaev-Viro construction \cite{TuraevViro, BarrettWestbury}. Again, it does not produce the most general oriented 3d TQFT\footnote{For example, Chern-Simons TQFTs for simple Lie groups do not arise from the Turaev-Viro construction.}, but it has the advantage that it can be easily extended to manifolds with boundaries. 

The starting point of the Turaev-Viro construction is a spherical fusion category $\cA$. It can be thought of as a categorification of a finite-dimensional semi-simple algebra $A$. A fusion category is a semi-simple rigid monoidal category with a finite number of simple objects. Let $I$ be the set of simple objects. The partition function is a sum over colorings of 1-simplices of the triangulation with elements of $I$.
If we denote such a coloring by $\phi$, we can write 
\begin{equation}
Z = \sum_\phi w[\phi] Z_\phi
\end{equation}
The weights $w[\phi]$ are a product over 1-simplces of an appropriate function of their color, times a product over all 0-simplices of an appropriate $\cA$-dependent constant.  

Each individual term $Z_\phi$ in the partition function can be thought of as the evaluation of a tensor network:
each face with edges of color $i$,$j$,$k$ is associated to a pair of vector spaces $V_{ijk}$, $V^*_{ijk}$, 
each one associated to a side of the face, and each tetrahedron $t$ to chosen element $F[\phi[t]]$ 
in the tensor product of the four vector spaces associated to its faces. The tensors are contracted together at the common faces. 
The partition function will be invariant under changes of triangulations if the $F[\phi[t]]$ tensors satisfy some basic axioms, such as 
the pentagon axiom, which guarantees invariance under a $2-3$ move. 

The same construction applied to a manifold with a boundary produces wave-functions, which can be reproduced as ground states of  
a Hamiltonian built from the same data \cite{LevinWen}.

The fermionic version of this construction is given in the literature in the ``Hamiltonian'' form, without a description 
of the corresponding partition partition function. In this section we will attempt to assemble the ingredients of the 
fermionic construction into a partition function. 

The starting point is a spherical super-fusion category $\cA$. As far as we know, this notion was first mentioned in \cite{fermionicTV} and appeared more recently in \cite{Walker} in the context of fermion condensation (see below). We will now describe what this means in a somewhat informal way, relegating the details to Appendix C. A super-fusion category has a finite number of simple objects $V_i$, $i\in I$ representing independent quasi-particle excitations. There are no nonzero morphisms between $V_i$ and $V_j$ if $i\neq j$, while the space of morphisms from $V_i$ to $V_i$ is a $\ZZ_2$-graded division algebra (i.e. every nonzero element has an inverse).  Unlike in the bosonic case, there are two possibilities for such a division algebra: $\CC$ or a Clifford algebra with one generator $\Cl(1)$.  Following \cite{fermionicTV} we will call the former kind of simple object a bosonic simple object, and the latter kind of object a Majorana simple object. Since $\Cl(1)\otimes\Cl(1)\simeq \Cl(2)$, the identity object is always bosonic. 

This can be explained in a different way. One can tensor any object of $\cA$  with a $\ZZ_2$-graded vector space. Tensoring with a bosonic vector space of dimension $N$ is equivalent to taking $N$ copies of the quasi-particle. Tensoring with $\CC^{0\vert 1}$ (a fermionic vector space of dimension $1$) has the physical meaning of attaching a fermion to the quasi-particle. The latter operation maps an object to an isomorphic one, but the obvious isomorphism is odd (fermionic). There may or may not be an even isomorphism as well. Now, suppose $V_i$ is a simple object. If $V_i$ is even-isomorphic to $\CC^{0\vert 1}\otimes V_i$, then $\Hom(V_i,V_i)\simeq \Cl(1)$, i.e. $V_i$ is Majorana. On the other hand, if $V_i$ is not even-isomorphic to $\CC^{0\vert 1}\otimes V_i$, then $\Hom(V_i,V_i)\simeq \CC$, i.e. $V_i$ is bosonic.

As usual, one can fuse quasi-particles, and the fusion rules are described by the way tensor products of simple objects decompose into sums of simple objects:
\begin{equation}\label {supertensorproduct}
V_i\otimes V_j\simeq \oplus_{k\in I} H_{ij}^k\otimes V_k.
\end{equation}
Here the ``coefficients'' $H_{ij}^k$ are finite-dimensional $\ZZ_2$-graded vector spaces. If some of the simple objects are Majorana, there is a subtlety: the absolute grading on some of $H_{ij}^k$ is not well-defined. For simplicity, we will assume that all simple objects are bosonic, so that all $H_{ij}^k$ have a well-defined grading.

The fusion of quasi-particles is associative, in the sense that for any $i,j,k\in I$ there is an even isomorphism
$$
\aa(i,j,k): (V_i\otimes V_j)\otimes V_k\ra V_i\otimes (V_j\otimes V_k).
$$
Expanding both sides in terms of simple objects, we deduce that $\aa(i,j,k)$ is determined by an even linear map of $\ZZ_2$-graded vector spaces
$$
F\begin{bmatrix}  i & j & l \\ k & m & n \end{bmatrix}: H^l_{ij}\otimes H^m_{lk}\ra H^m_{in}\otimes H^n_{jk}.
$$
This map is known as a 6j symbol. The collection of all 6j symbols satisfy an associativity constraint which ensures that the two ways of constructing an isomorphism from $(((V_i\otimes V_j)\otimes V_k)\otimes V_l)$ to $(V_i\otimes (V_j\otimes (V_k\otimes V_l)))$ are the same. To write down a concrete form for it, we need to choose a basis $e_{\alpha^k_{ij}},$ $\alpha^k_{ij}\in J^k_{ij},$ in each space $H^k_{ij}$. Let $\eps(\alpha^k_{ij})$ be the fermionic parity of the vector $e_{\alpha^k_{ij}}$. Then the associativity constraint (a.k.a. the fermionic pentagon equation) reads:
\begin{align}
&\sum_{t\in I} \Fj{i}{j}{m}{k}{n}{t}^{\alpha \beta}_{\eta \phi} \Fj{i}{t}{n}{l}{p}{s}^{\phi \chi}_{\kappa \gamma} \Fj{j}{k}{t}{l}{s}{q}^{\eta \kappa}_{\delta \phi} = \cr &=(-1)^{\epsilon(\alpha_{ij}^m)\epsilon(\delta_{kl}^q) }  \Fj{m}{k}{n}{l}{p}{q}^{\beta \chi}_{\delta \epsilon}\Fj{i}{j}{m}{q}{p}{s}^{\alpha \epsilon}_{\phi \gamma}.
\end{align}
Note the sign on the r.h.s. It reflects the fact that composition of morphisms in the symmetric tensor category of $\ZZ_2$-graded vector spaces is defined using the Koszul sign rule. 

A spherical structure on a super-fusion category assigns to every object $V$ its dual $V^*$, so that $V^{**}\simeq V$ and that for every two  objects $U,V$ we have isomorphisms
\begin{align}
\Hom&(U,V)\simeq \Hom(U\otimes V^*,1)\simeq \Hom(V^*\otimes U,1)\simeq\cr &\simeq \Hom(1,U^*\otimes V)\simeq \Hom(1,V\otimes U^*).
\end{align}
Let us denote by $V_\BI$ the dual of a simple object $V_i$. The spherical structure ensures that the spaces $H_{ijk}=H^\bk_{ij}$ are cyclically symmetric and that the space $H_{\bk\bj\BI}$ is  dual to $H_{ijk}$ for all $i,j,k\in I$.

In the fermionic  Turaev-Viro construction one considers colorings of 1-simplices of an oriented triangulated 3-manifold $X$ with objects $V_i$. We can choose a branching structure on the triangulation, so that the vertices of each 3-simplex and each 2-simplex have an order. To each 2-simplex one can attach a $\ZZ_2$-graded vector spaces $H_{ijk}$ and its dual, depending on the orientation. Next we would like to attach a 6j symbol, which is an element of a tensor product of four $\ZZ_2$-graded vector spaces, to a 3-simplex, and form the partition function by contracting these tensors along their shared 2-simplices. But there are two problems with this. First, the 6j symbol itself is an element of the tensor product with a particular ordering of $\ZZ_2$-graded factors $H_{ij}^k$ and their duals. Changing this order may change the sign of some components of the 6j symbol, but there is no natural order on the faces of a 3-simplex. Second, the contraction of two odd elements of dual $\ZZ_2$-graded vector spaces changes sign if one exchanges their order, but for a given 2-simplex in an oriented triangulation there is no preferred order of two 3-simplices sharing it, unless the 2-simplex itself is given an orientation. Such an orientation is induced by a branching structure, but this means that the partition function might depend on the branching structure. 

These choices are analogous to the choices  made in the construction of fermionic SPT phases: 
the order of the factors in the tensor product defining 6j symbols is analogous to the order of the Grassmann variables in $u[t]$
and the choice of order of the two sides of a 2-simplex is analogous to the choice of $d \theta d \bar \theta$ or $d \bar \theta d \theta$ in the fermionic measure.
One can make the analogy completely precise in the following way. First, we choose a basis
$$
e_{\alpha_{ijk}}, \alpha_{ijk}\in J_{ijk},
$$
in every space $H_{ijk}$.The weight of a particular coloring can be computed by summing over the variables $\alpha_{ijk}$ attached to every 2-simplex whose edges are colored by $i,j,k$. If we assume as before that the basis vectors have definite parity, each choice of $\alpha$ variables gives a $\ZZ_2$-valued 2-cochain. Further, since 6j symbols are even maps, this 2-cochain becomes a 2-cocycle $\beta_2$ when restricted to configurations of $\alpha$ variables for which all 6j symbols are nonzero. If we perform the summation over such variables first by fixing $\beta_2$, and then summing over all choices of $\beta_2$, then the Koszul sign is simply $\sigma_\Pi(\beta_2)$, where $\Pi$ denotes the ordering choices made in the definition of the partition function.

Thus the tensor network amplitude restricted to some fermionic parity sector $\beta_2$ 
can be decomposed as 
\begin{equation}
Z_\phi(\beta) = Z^b_\phi(\beta_2) \sigma_\Pi(\beta_2)
\end{equation}
where $Z^b_\phi$ is a ``bosonized'' amplitude computed without keeping track of 
Koszul signs in manipulating the vector spaces in the tensors, but rather pretends that all vector spaces are bosonic.

There is also a close analogy between the sign factor $(-1)^{\epsilon(\alpha_{ijm})\epsilon(\delta_{klq})}$ in the fermionic pentagon equation
and the twisted cocycle condition satisfied by the 3-cochain $\nu_3$ used in the construction of 3d fermionic SPT phases. 
Indeed, we can write the twisting factor as \footnote{In order to write this formula, we have assigned a specific order $01234$ to the vertices of 4-simplex, 
associated respectively to edges $imnp$, $ijs$, $kjmq$,$knl$, $lpqs$. 
This order is the one associated with the branching structure induced by the orientation of edges in the definition of the 6j symbols, such that a factor of $H_{ij}^k$ is associated to a positively oriented face, with 
edges $i$ and $j$ positively oriented and $k$ negatively oriented, and $(H_{ij}^k)^*$ is associated to a negatively oriented face, with 
edges $i$ and $j$ negatively oriented and $k$ positively oriented}
\begin{equation}
(-1)^{\epsilon(\alpha_{ijm})\epsilon(\delta_{klq}) } \equiv  (-1)^{\beta_2(012) \beta_2(234)} = (-1)^{[\beta_2\cup \beta_2](01234)} 
\end{equation}
This is precisely the cup square of $\beta_2$ evaluated on the 4-simplex associated to the 2-3 move.\footnote{We would like to thank Zhengcheng Gu for pointing it out.}
Notice that the fermionic pentagon relation involves a sum of terms with different fermionic parity in the internal faces, 
but the sign factor only involves the fermionic parity of the external faces. 

The fermionic pentagon relation is written in terms of the components of the tensors, and thus does not include any Koszul signs. 
The sign factor describes a change in the sign of the bosonic weight under a 2-3 move.
If we put the Koszul signs back in, the change of $\sigma_\Pi(\beta_2)$ under a 2-3 move will almost 
cancel the sign coming from the pentagon equation:  it will trade the cup square of $\beta_2$ with the usual linear term $\int w_2 \beta_2$
controlled by the canonical representative for $w_2$. Again, this linear term will provide an obstruction 
to assembling a well-defined partition function, unless the underlying manifold admits a spin structure. 

It should be now clear how to assemble a well-defined spin-structure-dependent partition sum:  
\begin{equation}
Z[\eta] = \sum_{\beta_2} \sum_\phi w[\phi] Z^b_\phi[\beta_2] z[X,\eta,\beta_{2}]
\end{equation}
where we decomposed the partition sum over eigenspaces of $\hat \beta_2$, denoted the ``bosonized'' 
partition sum in each sector as $Z^b_\phi[\beta_2]$ and combined the Koszul signs and 
the spin-structure-dependent correction into $z[X,\eta,\beta_{2}]$. The change of $z[X,\eta,\beta_{2}]$
under 2-3 moves will cancel against the sign factor in the pentagon relation, giving 
a well-defined object. 


Notice that $Z[\eta]$ is almost written as the partition function of a topological $\ZZ_2$ gauge theory with a 2-form gauge field $\beta_2$, 
except that $Z^b_\phi[\beta_2]$ is not well-defined on its own, as the $2-3$ move 
involves a summation over both bosonic and fermionic vector spaces when gluing the tetrahedra. 
Inspection of the $2-3$ move, though shows that the cocycle condition for $\beta_2$ 
almost fixes the Grassmann parity on internal faces, up to a binary choice, corresponding to 
a shift of $\beta_2$ by the coboundary of a cochain concentrated on the the ``t'' edge of the tetrahedra on the $3$ side
of the $2-3$ move. \footnote{Notice that because of that, the correction $\int_X \eta \cup \beta$ may add an extra 
relative sign between terms of a $2-3$ move, essentially controlled by the cup product 
between $w_2$ and a cochain concentrated on the ``t'' edge of the 4-symplex. 
That relative sign does not affect the $2-3$ move with the standard branching structure, where the ``t'' edge has vertices $(13)$
and does not contribute to a cup product. We expect that relative sign to be present in the pentagon relation for other choices of branching structure.}

That means that we can define a sensible bosonic partition function by summing only over $\beta_2$ which differ by an exact cocycle
\begin{equation}
\tilde Z^b[\beta_2] = \sum_{\lambda_1} \sum_\phi 2^{- \#(\mathrm{edges})}w[\phi] Z^b_\phi[\beta_2+ \delta \lambda_1] (-1)^{\int_X \beta_2 \cup \lambda_1 + \lambda_1 \cup \beta_2 + \lambda_1 \cup \delta  \lambda_1}
\end{equation}
and re-write the fermionic partition function as 
\begin{equation}
Z[\eta] = \sum_{\beta_2} \tilde Z^b[\beta_2]  z[X,\eta,\beta_{2}]
\end{equation}

Thus the fermionic Turaev-Viro theories can be obtained by gauging the diagonal $1$-form $\ZZ_2$ global symmetry of a product of appropriate bosonic theories 
and the spin-TFT $K_3$ defined by the partition function $z[X,\eta,\beta_{2}]$.

\section{Fermionic anyon condensation}\label{sec:condensation}
\subsection{Self-fermions and anomalous $1$-form $\ZZ_2$ global symmetry}
In this section we would like to argue that in $2+1$ dimensions, a bosonic TQFT equipped with a $1$-form 
$\ZZ_2$ global symmetry with the same 't Hooft anomaly as $K_3$ is the same as a TQFT which has in the spectrum of quasi-particles a self-fermion $\epsilon$ which fuses with itself to the identity. 

Our first observation is that a TQFT which has in the spectrum of quasi-particles 
a self-boson $b$ which fuses with itself to the identity is automatically equipped with a 
non-anomalous $1$-form $\ZZ_2$ global symmetry, whose generators are simply $b$ worldlines. 
The operation of gauging the $1$-form symmetry is the same as the operation of 
anyon condensation applied to $b$. 

If we are given any two TQFTs with self-fermions $\epsilon$ and $\epsilon'$ respectively which fuse with itself to the identity, 
the product of the two theories has a self-boson $\epsilon \epsilon'$. This shows that if we take $f$ and $f'$ worldlines
to be generators of $1$-form $\ZZ_2$ global symmetries, the two $\ZZ_2$ global symmetries 
will have the same $\ZZ_2$-valued 't Hooft anomaly.

That means we only need to compute the 't Hooft anomaly for some representative 
TQFT with a self fermion $\epsilon$ which fuses with itself to the identity. All other TQFTs with such a quasi-particle will have the same 't Hooft  anomaly. 

A simple choice of such a TQFT is the toric code, aka $\ZZ_2$ gauge theory. 
We can describe it by the discrete partition function 
\begin{equation}
(-1)^{\int_X a_1 \cup \delta b_1}
\end{equation}
with $a_1$ and $b_1$ being discrete cochains. 

The $e$,$m$ and $\epsilon$ quasi-particles correspond to Wilson loops for $a_1$, $b_1$ and $a_1 + b_1$ respectively. 
The former two quasi-particles are generators of two non-anomalous $1$-form $\ZZ_2$ symmetries.
\footnote{The two symmetries have a mixed anomaly.}
For example, we can couple the theory to a $\ZZ_2$ 2-cocycle $\beta^e_2$ as 
\begin{equation}
(-1)^{\int_X a_1 \cup \delta b_1 + a_1 \cup \beta^e_2}
\end{equation}
The action is invariant under gauge transformations $\beta_2^e + \delta \lambda_1^e$, accompanied by $b_1 \to b_1 + \lambda_1^e$. Similarly, we can couple the theory to a $\ZZ_2$ 2-cocycle $\beta^m_2$ as 
\begin{equation}
(-1)^{\int_X a_1 \cup \delta b_1 + \beta^m_2 \cup b_1}
\end{equation}
The action is invariant under gauge transformations $\beta_2^m + \delta \lambda_1^m$, accompanied by $a_1 \to a_1 + \lambda_1^m$.

On the other hand, the $1$-form $\ZZ_2$ symmetry associated to $\epsilon$ 
is encoded in the action 
\begin{equation}
(-1)^{\int_X a_1 \cup \delta b_1 + a_1 \cup \beta_2+ \beta_2 \cup b_1}
\end{equation}
Under gauge transformations $\beta_2 + \delta \lambda_1$, accompanied by $b_1 \to b_1 + \lambda_1$ and 
$a_1 \to a_1 + \lambda_1$, the action varies by 
\begin{equation}
(-1)^{\int_X \lambda_1 \cup \delta \lambda_1 + \lambda_1 \cup \beta_2+ \beta_2 \cup \lambda_1}
\end{equation}
This is precisely the anomalous variation for $K_3$! This verifies our claim. 

\subsection{Examples of fermionic anyon condensation}
We will come back to the toric code momentarily. 
Before that, we should give a simple examples of bosonic and fermionic theories related by 
fermionic anyon condensation. 

Consider first a $U(1)_{4k}$ Chern-Simons theory. The Wilson loop of charge $2k$ fuses with itself to the identity. 
If $k$ is even it is a self-boson $b$, while if $k$ is odd it is a self-fermion $\epsilon$. 
For even $k$, condensing $b$ leads to a $U(1)_k$ CS theory. We expect the same to occur for odd $k$, 
but then the resulting $U(1)_k$ CS theory is a spin-TQFT. This is consistent with the fact that this is a 
fermionic anyon condensation. 

It is interesting to observe that the toric code has a $\ZZ_2$ global symmetry which exchange the $e$ and 
$m$ particles. This symmetry is rather hard to make explicit in a concrete description of the model. 
For example, the action we wrote above is not invariant under a $\ZZ_2$ symmetry transformation
exchanging $a_1$ and $b_1$: 
\begin{equation}
(-1)^{\int_X a_1 \cup \delta b_1+ b_1 \cup \delta a_1} = (-1)^{\int_X \delta a_1 \cup_1 \delta b_1}
\end{equation}

Assuming that the $\ZZ_2$ global symmetry is not broken by the fermionic condensation of $\epsilon$, 
the resulting spin-TQFT should still enjoy the $\ZZ_2$ global symmetry. Indeed, there are good reasons to believe 
this is a fermionic SPT phase, the root fermionic SPT phase which is not captured by the super group cohomology 
classification. For example, we can consider the Hilbert space of the fermionic theory on a general Riemann surface 
(equipped with spin structure). Sectors with a non-trivial $2$-form gauge field flux are described by a 
Riemann surface with a single $\epsilon$ insertion, which do not contribute ground states. 
It is quite reasonable to assume that the $2^g$ ground states with no $\beta_2$ background flux
connection will be projected down to $2^g$ one-dimensional Hilbert spaces labelled by 
the choice of spin structure. 

A $\ZZ_2$-invariant boundary condition for the toric code has gapless modes: a non-chiral 2d Ising model 
coupled to the toric code in such a way that $e$ ends on the spin operator $\sigma(z, \bar z)$, $m$ on the dual spin operator 
$\mu(z,\bar z)$ and $\epsilon$ ends on chiral or anti-chiral fermion operators $\psi(z)$ and $\bar \psi(\bar z)$,
while the energy operator $\epsilon(z, \bar z)$ can exist without a quasi-particle world line ending on it.
The $\ZZ_2$ global symmetry is Kramer-Wannier duality. 

After condensing the $\epsilon$ quasi-particle, the chiral or anti-chiral fermion operators $\psi(z)$ and $\bar \psi(\bar z)$ 
in the boundary theory will be able to be inserted in correlation functions without a quasi-particle world line ending on it,
Thus the boundary theory reduces to a theory of free Majorana fermions, 
one chiral and the other anti-chiral, with the $\ZZ_2$ global symmetry acting only on one of the two 
fermions. This is an the expected feature of the root fermionic SPT phase with $\ZZ_2$ global symmetry. 

Another interesting example is the fermionic spin-TQFT ``s-Ising'' associated to the Ising 3d TQFT. The Ising TQFT has 
three quasi-particles: $1$, $\sigma$ and $\epsilon$. In the presence of a boundary, these quasi-particles end on 
the corresponding operators of a chiral Ising model. We expect that upon condensing the $\epsilon$ quasi-particle, 
the corresponding chiral operator on the boundary, which is a free fermion operator, will be free to appear by itself.
Thus the boundary theory should consist of a single chiral Majorana fermion, which can indeed be well-defined on a manifold 
equipped with a spin structure. 

It is also interesting to look at the Hilbert space of s-Ising on Riemann surfaces of various genus, 
equipped with a spin structure. On a torus, the Ising TQFT has three states. Upon gauging the 1-form symmetry, 
we expect them to be projected to three one-dimensional Hilbert spaces labelled by the three even spin structures on the torus. 
As they have no $\beta_2$ flux on the Riemann surface, they should have even fermion number. 
Upon gauging the 1-form symmetry we also gain twisted sectors with a non-zero flux of $\beta_2$ on the Riemann surface and odd fermion number, 
i.e. a single insertion of $\epsilon$. On the torus, that gives an extra state (labelled by a loop of $\sigma$ with an $\epsilon$ insertion),
which we expect to give a one-dimensional Hilbert spaces labelled by the single odd spin structure on the torus, 
of odd fermion number. 

On a genus $2$ Riemann surface the Ising TQFT has $10$ untwisted sectors, which should map to 
$10$ one-dimensional Hilbert spaces labelled by the $10$ even spin structures on the torus. 
We can also find $6$ states for the Ising TQFT on a genus two surface with an extra $\epsilon$ insertion,
which should map to $6$ one-dimensional Hilbert spaces labelled by the $6$ odd spin structures on the torus.

In general, we expect the s-Ising spin-TQFT to have one-dimensional spaces of ground states for 
every choice of spin structure on a Riemann surface, with even fermion number if the spin structure is even and odd 
if the spin structure is odd. In other words, if we compactly the s-Ising TQFT on a circle with Ramond boundary conditions, 
we should be left with a 2d theory which computes the Arf invariant of Riemann surfaces. 

It is somewhat surprising to observe that the s-Ising spin-TQFT has one-dimensional spaces of ground states, and yet 
supports chiral edge modes and would likely not be considered a fermionic SPT phase. Indeed, the standard classification 
of fermionic SPT phases in $2+1$ dimensions with fermion number symmetry only predicts no non-trivial theories.

\subsection{Bosonization in $2+1$ dimensions}
The operation of fermionic condensation on the toric code is somewhat intriguing from the perspective of defining a 
notion of bosonization in $2+1$ dimensions. Suppose that we were given a version of the toric code lattice Hamiltonian 
with the property that creating fermionic quasi-particle excitations could be created at much lower energy cost than 
the $e$ and $m$ quasi-particles. The Hilbert space of the low-energy theory will be the even part of a fermionic Fock space, 
but the interactions will not be completely local: fermion bilinear operators would still have to be represented by open string operators, 
corresponding to an $\epsilon$ worldline connecting the two fermionic creation/destruction operators. 

Suppose also that we could tensor that system with $K_3$ and gauge the diagonal 
$1$-form $\ZZ_2$ symmetry. Then the $\epsilon$ worldline string operators, dressed by the 
self-fermion quasiparticles of $K_3$, would become invisible in the gauged theory. 
This would liberate the endpoints of the open worldline string operators,
which would behave as fermionic creation or destruction operators. 
The Hilbert space of the gauged low-energy theory should become 
a fermionic Fock space, which one could use to simulate 
some general fermionic Hamiltonian in a local way. 

It would be interesting to make this recipe more precise. Ideally, one should find a lattice Hamiltonian 
description of the toric code which both allows the fermion quasiparticles to proliferate and 
which can be coupled appropriately to a $\beta_2$ cocycle.

\section*{Acknowledgments}
We thank Z.Gu and D. Freed for helpful discussions and J.  Morgan for educating one of us (AK) about spin structures and cobordism. 
The research of DG was supported by the
Perimeter Institute for Theoretical Physics. Research at Perimeter Institute is supported
by the Government of Canada through Industry Canada and by the Province of Ontario
through the Ministry of Economic Development and Innovation. The work of AK was
supported in part by the DOE grant DE-FG02-92ER40701 and by the Simons Foundation.
Opinions and conclusions expressed here are those of the authors and do not
necessarily reflect the views of funding agencies.

\appendix

\section{The Gu-Wen Grassmann integral in 2d as a quadratic refinement}
Notice that we can depict $\beta$ as a collection of closed, disjointed paths $C_a$ in the dual graph to the triangulation,
sequences of edges $e^{a}_i$ with $\beta(e^{a}_i)=1$ sharing a common triangle $t^a_{i \pm \frac{1}{2}}$ with the previous and next edges in the path. 
We can order the monomials in sequences along each path, and factorize the Grassmann integral as 
\begin{equation}
\sigma(\beta) = \prod_a \left[ \int \prod_i d \theta_{e^{a}_i} d \bar \theta_{e^{a}_i} \prod_i u(t^a_{i-\frac{1}{2}})\right]
\end{equation}
where we ordered the integrand in one direction and the measure in the opposite direction.
We assumed that $\epsilon[t,\beta]$ is $1$ unless a triangle is crossed by some path $C_a$. 

\begin{figure}
\begin{center}
\includegraphics[width=10cm]{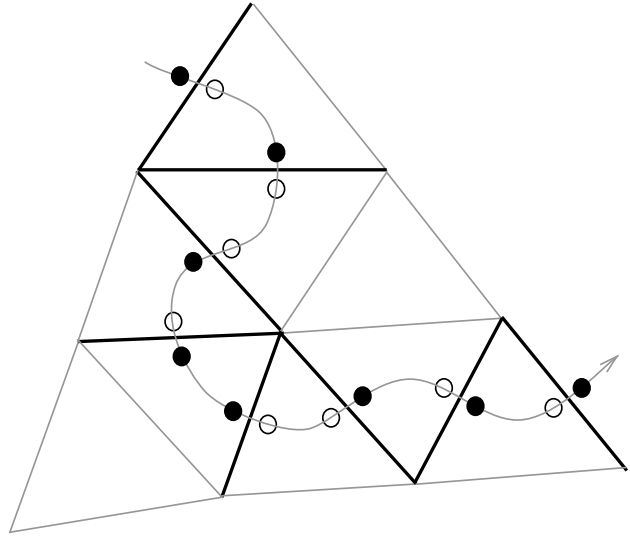}
\end{center}
\caption{
A collection of edges $e_i$ with $\beta(e_i)=1$, organized into a path in the dual graph to the triangulation. 
The Grassmann variables are encountered along the path in an order which differs by simple local permutations from the 
order they appear with in the measure and integrand of the Grassmann integral. }
\label{fig:var2}
\end{figure}

For each path $C_a$, denote as $c^a_i$ the sequence of Grassmann variables encountered along the path. 
We have 
\begin{equation}
\int \prod_i dc_{2i+1}^a dc_{2i}^a \prod_i c_{2i+1}^a c_{2i+2}=-1
\end{equation}
where $\ell_a$ is the length of $C_a$.  

The integration variables in $\prod_i d \theta_{e^{a}_i} d \bar \theta_{e^{a}_i}$ are encountered almost in the order 
$\prod_i dc_{2i+1}^a dc_{2i}^a$, up to permutations of two variables associated to the same edge. Similarly, 
the integration variables are encountered in the product  $\prod_i u(t^a_{i-\frac{1}{2}})$ almost in the same order 
as $\prod_i c_{2i+1}^a c_{2i+2}$, up to permutations of two variables associated to the same triangle. 

Thus the overall sign of the integral can be computed by multiplying a sign for each edge and each triangle along 
$C_a$:
\begin{equation}
\sigma(\beta) = \prod_a \left[ - \prod_i \epsilon_1[e^{a}_i] \prod_i \epsilon_2[t^a_{i-\frac{1}{2}}]\right]
\end{equation}
where the edge sign $\epsilon_1$ is $1$ if the edge orientation points to the right of $C_a$, $-1$ if it points to the left of $C_a$,
while the triangle sign $\epsilon_2$ is $1$ if both edges of the triangle crossed by $C_a$ point to the right of $C_a$, 
$-1$ if both point to the left, while if they point in opposite directions the sign is $1$ if $C_a$ is turning left at the triangle, $-1$ if turning right.

\begin{figure}
\begin{center}
\includegraphics[width=10cm]{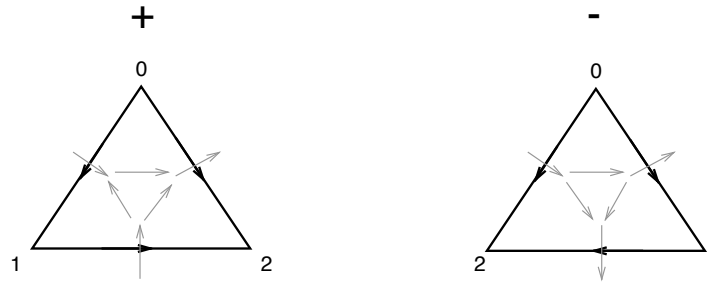}
\end{center}
\caption{
As a path $C$ proceeds along the triangulation, it accumulates a sign when crossing each edge, and a sign when passing through 
each triangle, depending on if the Grassmann variables it encounters are set in canonical order in the integral, or not. In this figure 
we indicate the directions along which a path receives a $+$ sign. The opposite directions produce a $-$ sign. }
\label{fig:pos}
\end{figure}

\begin{figure}
\begin{center}
\includegraphics[width=10cm]{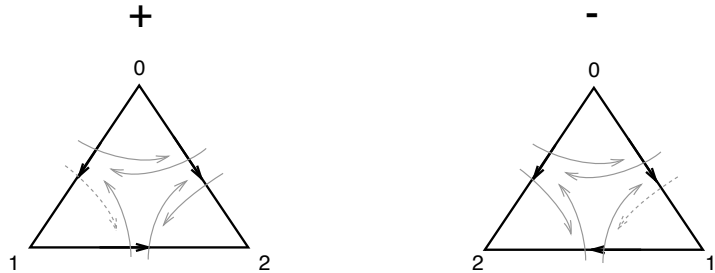}
\end{center}
\caption{
As a path $C$ proceeds along the triangulation from just before crossing an edge to just before crossing the next, it accumulates a sign as indicated in this figure: $+1$ in alms all cases, except the ones indicated by a dashed line.}
\label{fig:join}
\end{figure}

It is useful to combine the sign associated with crossing an edge to the sign associated to transversing the subsequent triangle. 
The resulting sign is almost always $+1$, except when we cross a $01$ edge and turn towards a $12$ edge. See Figures
\ref{fig:pos} and \ref{fig:join}.

\begin{figure}
\begin{center}
\includegraphics[width=10cm]{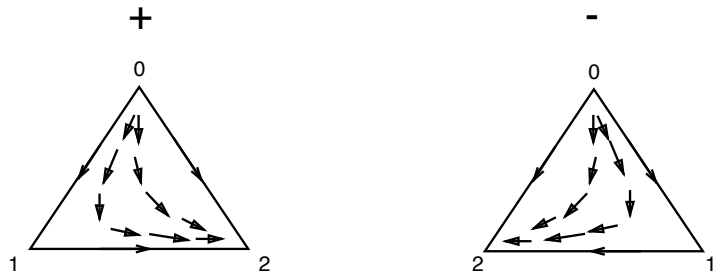}
\end{center}
\caption{
A useful reference vector field $V$ inside a triangle equipped with branching structure.}
\label{fig:flow}
\end{figure}

There is a simple interpretation of such sign. Consider a certain canonical vector field $V$ inside each triangle, nonzero away from the vertices and continuous across the edges, which flows out of the vertex $0$ and into the vertex $2$, turning around near the vertex $1$. This can be thought of as 
a continuation inside the triangle of the branching structure directions, as in Figure \ref{fig:flow}.

Given a smooth, non-self-intersecting path $C$ in the triangulation, which does not pass through the vertices, 
we can measure how many times the vector field $V$ restricted to $C$ winds around compared to the tangent 
vector of $C$. We can measure such winding number $w(V,C)$, modulo 2, by counting how many times the 
vector field $V$ becomes tangent to the path $C$. Inspection of the sign rule of \ref{fig:join}, shows that 
the overall sign accumulated by $C$ is $(-1)^{w(V,C)+1}$. 

Thus we conclude that 
\begin{equation}
\sigma(\beta) = \prod_a (-1)^{w(V,C_a)+1}
\end{equation}
if $\beta$ is represented by the collection of non-intersecting paths $C_a$. 

The sign $\sigma(\beta)$ does depend on the choice of cochain $\beta$ in a given cohomology class: 
if we sweep some path $C$ across a vertex $v$ of the triangulation, the winding number $w(V,C)$ will generically jump.
It is natural to expect, and it should be easy to show, that the jump modulo $2$ will be the same as 
the contribution of that vertex $v$ to the set $S$ of vertices which contribute to $Z_m$, 
i.e. $1$ plus the number of times $v$ is the vertex $1$ of a $-$ triangle. 

Indeed, that would insure that 
\begin{equation}
\sigma(\beta + \delta \lambda) = \sigma(\beta) \prod_{v \in S} (-1)^{\lambda(v)}
\end{equation}
so that the variation of $Z_\theta = \sigma(n)$ cancels against the variation of $Z_m$ under 
$m_1 \to m_1 + \lambda$. 

This fact is clearly true if the triangulation is a barycentric subdivision $BT$: 
the $0$ vertices are sources, the $1$ vertices simple saddles and the $2$ vertices are sinks. 
All these configurations are associated to a jump of the winding number by one unit. 

The combination 
\begin{equation}
(-1)^{q_E(\beta)} = \sigma(\beta) \prod_{e \in E} (-1)^{\beta(e)}
\end{equation}
is a known expression for the quadratic refinement of the intersection form 
on $\ZZ_2$ 1-cocycles on a surface \cite{Quadratic}. Besides transforming in the obvious way under changes of spin structure $E$, 
it satisfies 
\begin{equation}
q_E(\beta+ \beta') = q_E(\beta) + q_E(\beta') + \int \beta \cup \beta'
\end{equation}

Thus 
\begin{equation}
Z_n^E[BT] Z_\theta[BT] = (-1)^{q_E(n_1)}
\end{equation}

If we have a generic triangulation, we can still try to assemble a smooth vector field $V$ and 
look at the zeroes of $V$, with appropriate multiplicity, as a representative chain $S$ for $w_2$.
In general, we expect solutions of $\partial E = S$ to still describe a choice of spin structure 
on the discretized surface. One may interprete measuring the winding numbers $w(V,C)$ modulo 2, corrected by 
the pairing with $E$, as a way to build a spin bundle on the triangulated surface. 

\section{Higher cup products}\label{sec:higher}
As described in the introduction, higher cup products satisfy the 
recursive property 
\begin{equation} \label{eq:highercupbis}
A \cup_a B + B \cup_a A = \delta( A \cup_{a+1} B) + \delta A \cup_{a+1} B + A \cup_{a+1} \delta B
\end{equation}
with $\cup_0 \equiv \cup$. 

There is a rather explicit canonical formula for the higher cup products \cite{Higher}: 
\begin{align} \label{eq:higher}
&[A_p \cup_a B_q](0, \cdots, p+q) = \cr & \sum_{i_0<\cdots<i_a} A(0,\cdots i_0, i_1, \cdots i_2, i_3, \cdots) B(i_0, \cdots, i_1, i_2, \cdots, i_3, i_4, \cdots)
\end{align}
where the sequences of arguments of $A_p$ and $B_q$ must agree with the degree $p$,$q$ of $A_p$ and $B_q$. 

For example, we recover the standard cup product: 
\begin{equation} \label{eq:cup}
[A_p \cup B_q](0, \cdots, p+q) =  A(0,\cdots p) B(p, \cdots,p+q)
\end{equation}
The next cup products are 
\begin{equation} \label{eq:cupone}
[A_p \cup_1 B_q](0, \cdots, p+q-1) = \sum_{i_0} A(0,\cdots i_0, i_0+q, p+q-1) B(i_0, \cdots, i_0+q)
\end{equation}
and
\begin{align} \label{eq:cuptwo}
[A_p \cup_2 B_q]&(0, \cdots, p+q-2)= \cr  = \sum_{i_0<i_1} & A(0,\cdots i_0, i_1, \cdots p+i_1-i_0-1) \cdot\cr & \cdot  B(i_0, \cdots, i_1, p+i_1-i_0-1, \cdots, p+q-2)
\end{align}
Etcetera. 

\section{Super-fusion categories}

A category $\cC$ enriched over $\Vect_\kk$ (the category of finite-dimensional vector spaces over a field $\kk$) is a category whose morphism sets $\Hom(A,B)$ are vector spaces over $\kk$, and the composition of morphisms is bilinear. It can be thought of as a generalization of the notion of an algebra over a field (``algebra with many objects''). Similarly, a category $\cC$ enriched over $\sVect_\kk$ (the category of $\ZZ_2$-graded vector spaces over a field $\kk$)  is a category whose morphism sets $\Hom(A,B)$ are $\ZZ_2$-graded vector spaces over $\kk$, and the composition of morphisms is bilinear and is compatible with the grading. Such a catregory $\cC$ can be thought of a generalization of the notion of a $\ZZ_2$-graded algebra over a field. In what follows we will suppress $\kk$ (in physical applications one typically has $\kk=\CC$).

An important difference between ordinary algebras and $\ZZ_2$-graded algebras is the way the tensor product of algebras is defined. For ordinary algebras, the product of algebras $C$ and $D$ is an algebra whose underlying set is $C\otimes D$ and the multiplication is defined by
$$
(c\otimes d)\cdot(c'\otimes d')=cc'\otimes dd',\quad c,c'\in C,\ d, d'\in D.
$$
On the other hand, for $\ZZ_2$-graded algebras one defines the multiplication by
$$
(c\otimes d)\cdot(c'\otimes d')=(-1)^{\eps(c')\eps(d)} (c\cdot c')\otimes (d\cdot d'),\quad c,c'\in C,\ d, d'\in D,
$$
where $\eps(a)\in\ZZ_2$ is the fermionic parity of $a\in C$.
This is the Koszul sign rule. This has an analog for categories: the Cartesian product of categories over $\sVect$ is defined differently from the Cartesian product of categories over $\Vect$. In the former case, if $C,C', C''$ and $D,D',D''$ are objects of $\cC$ and $\cD$ respectively, the composition of morphisms in $\cC\times\cD$ is defined as follows:
\begin{multline}
(\alpha\otimes\beta) \cdot (\gamma\otimes\delta)=(-1)^{\eps(\gamma)\eps(\beta)} (\alpha\cdot\gamma)\otimes (\beta\cdot\delta),\\
 \alpha\in \Hom_\cC(C',C''),\gamma\in\Hom_\cC(C,C'), \\ \beta\in \Hom_\cD(D',D''),\delta\in \Hom_\cD(D,D').
\end{multline}
while in the latter case the sign factor on the r.h.s. is absent. \footnote{More generally, one can consider categories enriched over a symmetric tensor category. The definition of Cartesian product for such categories depends on the symmetric tensor category.}

This in turn has consequences for how a tensor product on a category is defined. Both in the ordinary and the $\ZZ_2$-graded case, a tensor product on $\cC$ is a functor $m$ from $\cC\times \cC$ to $\cC$, $m: (X,Y)\mapsto X\otimes Y$, $\forall X,Y\in \Ob(\cC)$, together with an associator. The associator is a natural isomorphism between two functors from $\cC\times\cC\times\cC$ to $\cC$ built from $m$. The first one is $m(m(-,-),-)$ and the second one is $m(-,m(-,-))$. The associator must satisfy the pentagon equation. Since the definition of the Cartesian product depends on whether we consider categories over $\Vect$ or $\sVect$, so does the definition of the tensor structure. 

Let us specialize to the case of (multi)-fusion categories. Multi-fusion categories are  rigid (and in particular semi-simple) monoidal categories (categories with a tensor product and a unit object) over $\Vect$ or $\sVect$ which have finitely many isomorphism classes of simple objects. To emphasize the difference between the two cases, we will refer to multi-fusion categories over $\sVect$ as super-multi-fusion categories, while multi-fusion categories over $\Vect$ will be simply called multi-fusion categories. While multi-fusion categories are module categories over $\Vect$, super-multi-fusion categories are module categories over $\sVect$. 

Let $\cC$ be a super-multi-fusion category. For every object $X\in \Ob(\cC)$ we have a $\ZZ_2$-graded algebra $\Hom_\cC(X,X)$. If $X$ is simple, this algebra must be a division algebra, therefore it is isomorphic either to $\CC$ or $\Cl(1)$. In the former case we will say that $X$ is a bosonic simple object, while in the latter case we will say that $X$ is a Majorana simple object. 

A super-fusion category is a super-multi-fusion category whose unit object $\bone$ is simple. Since $\bone\otimes\bone\simeq\bone$, this implies that $\Hom(\bone,\bone)=\CC$, i.e. $\bone$ is a bosonic object.  

Since a super-fusion category is rigid, for every object $V$ we have a dual object $V^*$, and even morphisms ${\rm ev}_L:V^*\otimes V\ra\bone$, ${\rm coev}_L:\bone \ra V\otimes V^*,$ ${\rm ev}_R: V\otimes V^*\ra\bone$ and ${\rm coev}_R: \bone\ra V^*\otimes V$ satisfying the usual identities. One can show \cite{DGNO} that $V^{**}$ is isomorphic to $V$. A pivotal structure on a super-fusion category is a choice of such isomorphisms for all $V$ in a way compatible with the tensor product. Given a pivotal structure, one can define
the left and right dimensions of $V$ by composing ${\rm coev}_L$ and ${\rm ev}_R$ or ${\rm coev}_R$ and ${\rm ev}_L$. The left and right dimensions are complex numbers. If the left and right dimensions are equal for all $V$, one says that the pivotal structure is spherical. A spherical super-fusion category is a super-fusion category equipped with a spherical pivotal structure.


\begin{thebibliography}{99}

\bibitem{ChenGuWen} X.~Chen, Z.~-C.~Gu,  and X.~-G.~Wen, ``Local unitary transformations, long-range quantum entanglement, wave function renormalization, and topological order,'' Phys. Rev. {\bf B 82}, 155138 (2010).

\bibitem{Kitaevtalk} A. Kitaev, talk at the IPAM workshop ``Symmetry and Topology in Quantum Matter,'' Jan 26-30, 2015. 

\bibitem{Witten:index} E.~Witten, ``Supersymmetric index of three-dimensional gauge theory,''
  In *Shifman, M.A. (ed.): The many faces of the superworld* 156-184
  [hep-th/9903005].
  
\bibitem{Wen:mono} X.~G.~Wen, ``Quantum field theory of many-body systems: From the origin of sound to an origin of light and electrons,''
  Oxford, UK: Univ. Pr. (2004).
  
\bibitem{ChenGuWen2} X.~Chen, Z.~-C.~Gu,  and X.~-G.~Wen, ``Classification of gapped symmetric phases in 1D spin systems,'' Phys. Rev. {\bf 83}, 035107 (2011).

\bibitem{KitaevFidkowski} L.~Fidkowski,  A.~Kitaev,` `Topological phases of fermions in one dimension,'' Phys. Rev. {\bf B 83}, 075103 (2011). 



\bibitem{GuWen} Z-C. Gu and X-G. Wen, ``Symmetry-protected topological orders for interacting fermions: Fermionic topological nonlinear $\sigma$-models and a special group supercohomology theory,''  Phys. Rev. {\bf B 90}, 115141 (2014) [arXiv:1201.2648 [cond-mat]].

\bibitem{fermionictoriccode} Z-C. Gu, Z. Wang, X-G. Wen, ``Lattice model for fermionic toric code,'' Phys. Rev. {\bf B 90}, 085140 (2014) [arXiv:1309.7032 [cond-mat]].
 
\bibitem{fermionicTV} Z-C. Gu, Z. Wang, X-G. Wen,  ``A classification of 2D fermionic and bosonic topological orders,'' Phys. Rev. {\bf B 91}, 125149 (2015) [arXiv:1010.1517 [cond-mat]].

\bibitem{LevinWen} M.~A.~Levin and X.~G.~Wen, ``String net condensation: A Physical mechanism for topological phases,''
  Phys.\ Rev.\ B {\bf 71}, 045110 (2005) [cond-mat/0404617].

\bibitem{Kap:cobord} A.~Kapustin, ``Symmetry Protected Topological Phases, Anomalies, and Cobordisms: Beyond Group Cohomology,''
  arXiv:1403.1467 [cond-mat.str-el].
  
\bibitem{FSPT} A.~Kapustin, R.~Thorngren, A.~Turzillo and Z.~Wang, ``Fermionic Symmetry Protected Topological Phases and Cobordisms,''
  arXiv:1406.7329 [cond-mat.str-el].
  
\bibitem{LevinGu} Z-C. Gu and M. Levin, ``The effect of interactions on 2D fermionic symmetry-protected topological phases with $Z_2$ symmetry,'' Phys. Rev. {\bf B 89}, 201113 (2014) [arXiv:1304.4569[cond-mat]].

\bibitem{Stong} R. E. Stong, ``Notes on cobordism theory,'' Princeton University Press (1968).

\bibitem{Atiyah} M. F. Atiyah, ``Riemann surfaces and spin structures,'' Ann. Sci. Ecole Norm. Sup. {\bf 4}, 47-62 (1971).

\bibitem{Johnson} D. Johnson, ``Spin structures and quadratic forms on surfaces,'' J. London Math. Soc. {\bf 22}, 365-373 (1980). 
  
\bibitem{latticeTQFT} M.~Fukuma, S.~Hosono and H.~Kawai, ``Lattice topological field theory in two-dimensions,'' Commun.\ Math.\ Phys.\  {\bf 161}, 157 (1994) [hep-th/9212154].

\bibitem{BP} C.~Bachas and P.~M.~S.~Petropoulos, ``Topological models on the lattice and a remark on string theory cloning,''
 Commun.\ Math.\ Phys.\  {\bf 152}, 191 (1993) [hep-th/9205031].

\bibitem{TuraevViro} V. Turaev and O. Ya. Viro, ``State sum invariants of 3-manifolds and quantum 6j symbols,'' Topology {\bf 31} (1992), no. 4, 865Ð902.

\bibitem{BarrettWestbury} J. Barrett and B. Westbury, ``Invariants of piecewise-linear 3-manifolds,'' Trans. Amer. Math. Soc. {\bf 348} (1996), no. 10, 3997Ð4022.

\bibitem{DGNO} V. Drinfeld, S. Gelaki, D. Nikshych, V. Ostrik, ``On braided fusion categories,'' Selecta Mathematica {\bf 16},  no. 1, 1-119. 

\bibitem{MooreSegal} G.~W.~Moore and G.~Segal, ``D-branes and K-theory in 2D topological field theory,'' hep-th/0609042.

\bibitem{Quadratic} D.Cimasoni and N. Reshetikhin, ``Dimers on Surface Graphs and Spin Structures,''. Comm. Math. Phys. {\bf 275} (2007), no. 1, 187-208 math-ph/0608070.

\bibitem{Higher} N. Steenrod, ÒProducts of cocycles and extensions of mappings,Ó Ann. of Math. (2) 48, 290 (1947).

\bibitem{Walker} K. Walker, talk at the IPAM workshop ``Symmetry and Topology in Quantum Matter,'' Jan. 26-30, 2015.


\bibitem{BarrettTavares} J. W. Barrett and S. O. G. Tavares, ``Two-dimensional state sum models  and spin structures,'' arXiv:1312.7561 [math-ph]. 

\bibitem{NovakRunkel} S. Novak and I. Runkel, ``State-sum constructions of two-dimensional topological quantum field theories on spin surfaces,'' arXiv:1402.2839 [math.QA].

\end{thebibliography}

\end{document}